\documentclass[twocolumn,tighten,times]{aastex63}

\usepackage{newtxtext,newtxmath}
\usepackage{graphicx}	
\usepackage{subfigure}
\usepackage{amsmath}	
\usepackage{amssymb}	
\usepackage{rotating}
\usepackage{longtable}
\usepackage{threeparttable}
\usepackage{float}

\shorttitle{SOFIA Atlas}
\shortauthors{Fuller et al.}

\begin{document}

\title{The Galaxy Activity, Torus, and Outflow Survey (GATOS). VII. The 20-214 $\mu$m imaging atlas of active galactic nuclei using SOFIA}

\author[0000-0003-4809-6147]{Lindsay Fuller}
\affiliation{University of Texas at San Antonio, One UTSA Circle, San Antonio, TX, 78249, USA}

\author[0000-0001-5357-6538]{Enrique Lopez-Rodriguez}
\affiliation{Department of Physics \& Astronomy, University of South Carolina, Columbia, SC 29208, USA}
\affiliation{Kavli Institute for Particle Astrophysics \& Cosmology (KIPAC), Stanford University, Stanford, CA 94305, USA}

\author[0000-0002-9627-5281]{Ismael Garc\'ia-Bernete}
\affiliation{Department of Physics, University of Oxford, Keble Road, Oxford OX1 3RH, UK}

\author[0000-0001-8353-649X]{Cristina Ramos Almeida}
\affiliation{Instituto de Astrof\'{i}sica de Canarias, Calle V\'{i}a L\'{a}ctea, s/n, E-38205 La Laguna, Tenerife, Spain}
\affiliation{Departamento de Astrofísica, Universidad de La Laguna, E-38206 La Laguna, Tenerife, Spain}

\author[0000-0001-6794-2519]{Almudena Alonso-Herrero}
\affiliation{Centro de Astrobiolog\'ia (CAB), CSIC-INTA, Camino Bajo del Castillo s/n, E-28692, Villanueva de la Ca\~nada, Madrid, Spain}

\author[0000-0001-7827-5758]{Chris Packham}
\affiliation{University of Texas at San Antonio, One UTSA Circle, San Antonio, TX, 78249, USA}
\affiliation{National Astronomical Observatory of Japan, National Institutes of Natural Sciences (NINS), 2-21-1 Osawa, Mitaka, Tokyo 181-8588, Japan}

\author[0000-0003-4937-9077]{Lulu Zhang} 
\affiliation{University of Texas at San Antonio, One UTSA Circle, San Antonio, TX, 78249, USA}

\author[0000-0003-4975-2046]{Mason Leist} 
\affiliation{University of Texas at San Antonio, One UTSA Circle, San Antonio, TX, 78249, USA}

\author[0000-0003-4209-639X]{Nancy Levenson} 
\affiliation{Space Telescope Science Institute, 3700 San Martin Drive, Baltimore, MD 21218, USA}

\author[0000-0001-6186-8792]{Masa Imanishi} 
\affiliation{National Astronomical Observatory of Japan, National Institutes of Natural Sciences (NINS), 2-21-1 Osawa, Mitaka, Tokyo 181-8588, Japan}

\author[0000-0002-6353-1111]{Sebastian Hoenig}
\affiliation{School of Physics \& Astronomy, University of Southampton, Southampton SO17 1BJ, UK}

\author[0000-0001-5146-8330]{Marko Stalevski} 
\affiliation{Astronomical Observatory, Volgina 7, 11060 Belgrade, Serbia}
\affiliation{Sterrenkundig Observatorium, Universiteit Gent, Krijgslaan 281-S9, Gent B-9000, Belgium}

\author[0000-0001-5231-2645]{Claudio Ricci} 
\affiliation{Núcleo de Astronomía de la Facultad de Ingeniería, Universidad Diego Portales, Av. Ej\'{e}rcito Libertador 441, Santiago, Chile}
\affiliation{Kavli Institute for Astronomy and Astrophysics, Peking University, Beijing 100871, People’s Republic of China}

\author[0000-0002-4457-5733]{Erin Hicks} 
 \affiliation{Department of Physics and Astronomy, University of Alaska Anchorage, Anchorage, AK 99508-4664, USA}

\author[0000-0001-9791-4228]{Enrica Bellocchi} 
\affiliation{Departmento de F\'{i}sica de la Tierra y Astrof\'{i}sica, Fac. de CC F\'isicas, Universidad Complutense de Madrid, E-28040 Madrid, Spain}
\affiliation{Instituto de F\'{i}sica de Part\'iculas y del Cosmos IPARCOS, Fac. CC F\'isicas, Universidad Complutense de Madrid, E-28040 Madrid, Spain}

\author[0000-0003-2658-7893]{Francoise Combes} 
\affiliation{LERMA, Observatoire de Paris, Coll\'{e}ge de France, PSL University, CNRS, Sorbonne University, Paris, France}

\author[0000-0003-4949-7217]{Ric Davies} 
\affiliation{Max Planck Institut f\"ur Extraterrestrische Physik, Giessenbachstrasse 1, D-85748 Garching bei M\"unchen, Germany}

\author[0000-0003-0444-6897]{Santiago García Burillo} 
\affiliation{Observatorio de Madrid, OAN-IGN, Alfonso XII, 3, E-28014 Madrid, Spain}

\author[0000-0002-2356-8358]{Omaira Gonz\'alez Martín} 
\affiliation{Instituto de Radioastronom\'ia y Astrofísica (IRyA), Universidad Nacional Aut\'onoma de M\'exico, Antigua Carretera a P\a'tzcuaro \#8701, ExHda. San Jos\'e de la Huerta, Morelia, Michoac\'an, C.P. 58089, Mexico}

\author[0000-0001-9452-0813]{Takuma Izumi} 
\affiliation{National Astronomical Observatory of Japan, National Institutes of Natural Sciences (NINS), 2-21-1 Osawa, Mitaka, Tokyo 181-8588, Japan}

\author[0000-0002-0690-8824]{Alvaro Labiano} 
\affiliation{Telespazio UK for the European Space Agency, ESAC, Camino Bajo del Castillo s/n, E-28692 Villanueva de la Ca\~nada, Spain}

\author[0000-0002-4005-9619]{Miguel Pereira Santaella} 
\affiliation{Instituto de F\'isica Fundamental, CSIC, Calle Serrano 123, E-28006 Madrid, Spain}

\author[0000-0001-6854-7545]{Dimitra Rigopoulou} 
\affiliation{Department of Physics, University of Oxford, Keble Road, Oxford OX1 3RH, UK}
\affiliation{School of Sciences, European University Cyprus, Diogenes street, Engomi, 1516 Nicosia, Cyprus}

\author[0000-0002-0001-3587]{David Rosario} 
\affiliation{School of Mathematics, Statistics and Physics, Newcastle University, Newcastle upon Tyne NE1 7RU, UK}

\author{Daniel Rouan}
\affiliation{LESIA, Observatoire de Paris, Universit\'{e} PSL, CNRS, Sorbonne Universit\'{e}, Sorbonne Paris Cite\'{e}, 5 place Jules Janssen, F-92195 Meudon, France}

\author[0000-0002-2125-4670]{Taro Shimizu} 
\affiliation{Max Planck Institut f\"ur Extraterrestrische Physik, Giessenbachstrasse 1, D-85748 Garching bei M\"unchen, Germany}

\author{Martin Ward}
\affiliation{Centre for Extragalactic Astronomy, Department of Physics, Durham University, South Road, Durham DH1 3LE, UK}

\begin{abstract}

We present a $19.7-214~\mu$m imaging atlas of local ($4-181$ Mpc; median 43 Mpc) active galactic nuclei (AGN) observed with FORCAST and HAWC+ on board the SOFIA telescope with angular resolutions $\sim$ $3\arcsec - 20\arcsec$. This atlas comprises $22$ Seyferts ($17$ Type 2 and $5$ Type 1) with a total of $69$ images, $41$ of which have not been previously published. The AGN span a range of luminosities of $\log_{10} (L_{\rm bol}[\rm{erg/s}]) = [42,46]$ with a median of $\log_{10} (L_{\rm bol}[\rm{erg/s}]) = 44.1\pm1.0$. We provide total fluxes of our sample using aperture photometry for point source objects and a 2-D Gaussian fitting for objects with extended host galaxy emission, which was used to estimate the unresolved nuclear component. Most galaxies in our sample are point-like sources, however, four sources (Centaurus A, Circinus, NGC 1068, and NGC 4388) show extended emission in all wavelengths. The $30-40$ $\mu$m extended emission in NGC 4388 is coincident with the narrow line region at PA $\sim50^{\circ}$, while the dusty extension at longer wavelengths arises from the host galaxy at PA $\sim 90^{\circ}$. Our new observations allow us to construct the best sampled spectral energy distributions (SEDs) available between 30 - 500 $\mu$m for  a sample of nearby AGN. We estimate that the average peak wavelength of the nuclear SEDs is $\sim40~\mu$m in $\nu F_{\nu}$, which we associate with an unresolved extended dusty region heated by the AGN.  

\end{abstract}


\section{Introduction}
\label{sec:intro}

There is clear evidence that a considerable amount of dust in the vicinity of supermassive black holes (SMBHs) in active galaxies obscures the central engine (i.e., accretion disk and SMBH) in some lines of sight. Through spectropolarimetric observations of NGC 1068, \citet{Antonucci1985} showed that its optical polarized spectrum contained broad optical polarized emission lines not originally observed by direct total intensity observations. It was subsequently presumed that an optically and geometrically thick dusty structure (`torus') blocked the central engine in some lines of sight \citep{Antonucci1993,Urry1995}. Under this unified scheme a Type 1 AGN is seen face-on and shows broadened optical lines, while in Type 2 AGN the broadened lines are obscured. This model also predicts that broad silicate features at 10 and 18 $\mu$m will be seen in emission in Type 1 and in absorption in Type 2. However, silicate emission can be seen in emission in some Type 2 AGN, while absorption can be seen in some Type 1 \cite[e.g.,][]{Hatziminaoglou2015}. This and other observational features are explained by the inhomogeneous nature of the torus. Clumpy torus models \citep{Nenkova2008a,Nenkova2008b} predict shallower silicate features, more similar infrared spectral energy distributions (SEDs) between Type 1 and Type 2 AGN, etc. \citep[see][for a review]{RA2017}. 

A region of narrow forbidden line emission extends above and below the midplane of the dusty torus structure out to several kpc scales. Recent sub-arcsecond interferometric imaging observations have shown a dust component at pc-scales co-spatial with the base of the narrow line region \citep[NLR;][]{Honig2012,Tristram2014,LG2014,LG2016,Burtscher2013,GR2022,Isbell2022}. This dusty structure is interpreted as part of a dusty wind launched from the inner hot part of the torus driven by radiation pressure at pc-scales \citep{Honig2019}, but generated by a magnetohydrodynamical wind at sub-pc scales \citep[e.g.,][]{Emmering1992,ELR2015,Takasao2022,ELR2023}. This extended dusty structure has been resolved in a nearby galaxy, ESO 418-G14, using mid-infrared (MIR) images with JWST/MIRI finding that the dust is primarily heated by the AGN and/or radiative jet-induced shocks in the NLR rather than a wind \citep{Haidar2024}.

ALMA observations provide observational support for a dusty torus+outflow scenario. Emission from the nucleus of NGC 1068 was mapped with a resolution of $\sim 4$ pc, resolving a $7-10$ pc diameter disk interpreted as the sub-mm counterpart of the torus \citep{GB2016}. Rotation of the compact emission was detected in HCN J=3-2 and HCO+ J=3-2 \citep{Imanishi2018,Imanishi2020}. A molecular outflowing wind co-spatial with the dusty and molecular torus was also observed \citep{GB2019}. 
\citet{AH2018} interpreted the measured nuclear ($10-20$ pc) CO(2-1) emission in NGC 5643 as a nuclear molecular gas component of the torus that is likely collimating the ionization cone. Conditions favorable for launching a cold and molecular wind likely depend on Eddington ratio and nuclear hydrogen column densities  \citep[e.g.,][]{Venanzi2020,GB2021,AH2021,GBernete2022}.

This pc-scale dusty component is possibly associated with larger scale MIR emission detected out to $100$s pc scales. In the case of Circinus, high angular resolution MIR imaging, optical polarimetry and integral field spectra, coupled with state-of-the-art radiative transfer simulations, provide evidence that extended dust emission from pc to tens of pc scales in this object is a result of a hollow dusty cone illuminated by a tilted accretion disk \citep{Stalevski2017,Stalevski2019,Stalevski2023,Kakkad2023}. MIR extended emission out to 1" ($\sim$ 75 pc) was clearly detected in NGC 1068 by \citet{Bock2000}. Later $10.8$ and $18.2$ $\mu$m emission extending $3\farcs5$ ($\sim200$ pc) across NGC 4151 was also attributed to dust in the NLR heated by the central engine \citep{Radomski2003}. Likewise, at similar wavelengths, extended emission in $18$ AGN at distances out to hundreds of parsecs was detected \citep{Asmus2016,GBernete2016,Asmus2019}. Using the $37.1$ $\mu$m filter on SOFIA/FORCAST and thanks to the increase in angular resolution compared with $Spitzer$, extended dust emission in Mrk 3, NGC 4151, and NGC 4388 was found on $\sim100$s pc-scales \citep{F19} coincident with the NLR and radio axis. This emission may be due to dust along the wall of ionization cones \citep{Mason2009} or a dusty NLR \citep{Mor2009,Mor2012}.

In this manuscript we present an imaging atlas of $22$ local (D $=4 - 181$ Mpc; median 42.8 Mpc) AGN obtained using the FORCAST and HAWC+ instruments on the 2.7-m SOFIA telescope in the wavelength range $20 - 214~\mu$m. Most of these datasets are unpublished or dispersed throughout the literature. We provide a mid- to far-IR imaging atlas at angular scales of $\sim 3 - 20\arcsec$. At these scales, contribution from several dust sources is expected and we expect to disentangle the emission sources in a future study. Instead, here we aim to determine whether these objects are extended or not, and at what wavelengths within the resolution of the SOFIA telescope. We also explore the wavelength of turnover in the SED. This atlas is complementary to JWST observations up to $\sim25~\mu$m and archival \textit{Herschel} data ($ 70-500~\mu$m). 

The manuscript is organzed as follows. Section \ref{sec:OBS} describes the observations and AGN sample definition; Section \ref{sec:images} shows the new IR images; Section \ref{sec:analysis} contains details of the imaging analysis and Section \ref{sec:SEDs} shows the resulting SEDS; we  present results about the data in Section \ref{conclusions}.


\section{AGN Sample and Observation Data}
\label{sec:OBS}

\subsection{Sample Selection} 
\label{subsec:AGNsample}

This imaging atlas was drawn from the ongoing AGN survey performed by the Galactic Activity, Torus, and Outflow Survey \citep[GATOS;][]{GB2021,AH2021,IGB2024}. GATOS\footnote{GATOS website: \url{https://gatos.myportfolio.com/}} aims to characterize the dynamics and composition of the dusty and molecular torus and multi-phase outflows in AGN. The GATOS parent sample is selected from the 70 Month \textit{Swift}/BAT AGN catalog, which is flux limited in the ultra-hard $14-195$ keV X-rays band \citep{Baumgartner2013}. 

In the initial study of AGN using SOFIA observations \citep{F16}, sources from the GATOS survey were selected based on the criteria that the galaxies had been previously studied using \textsc{Clumpy} \citep{Nenkova2008a,Nenkova2008b} torus models and were well-sampled in the $1 - 18~\mu$m regime \citep{RA2009,RA2011,AH2011}. The study of the 11 objects included the $31.5~\mu$m photometry in the SEDs and found that including the 31.5 $\mu$m photometry reduces the number of \textsc{Clumpy} torus models that are compatible with the data and modifies the model output for the torus outer radius. \citet{F19} further extended the wavelength range of a subset of $7$ AGN SEDs to 37.1 $\mu$m. They subsequently found extended emission in the PSF-subtracted images of Mrk 3, NGC 4151, and NGC 4388 that is coincident with the radio axis and NLR. In a separate study, \citet{ELR2018} modeled the torus of NGC~1068 using $\sim 20 - 53~\mu$m FORCAST and HAWC+ observations. They showed that the peak wavelength range of emission from the torus is $\sim$ 30 - 40 $\mu$m with a characteristic temperature 70 - 100 K. The use of observations > 30 $\mu$m in that study from SOFIA and ALMA highlights the importance of longer wavelength observations to put constraints on MIR emission sources. Based on these results, we extend the wavelength range in objects previously observed, and also expand the number of AGN observed. 

We present the complete imaging atlas of $22$ Seyferts observed by SOFIA in the wavelength range $19.7 - 214~\mu$m using FORCAST and HAWC+. The final set of observations presented here was part of a multi-year AGN survey over several observing SOFIA cycles (Proposal IDs: 02\_0035, 04\_0048, 06\_0066, 08\_0014; PI: Lopez-Rodriguez; 70\_0400 PI: Herter). The SOFIA atlas of AGN in the far-IR (FIR) is a flux-limited sample of nearby, bright, and well-studied AGN. All objects have a point source flux of $>200$ mJy at $31.5~\mu$m, which ensures that each band can be observed within $1$ hr of on-source time with a signal to noise ratio $>10$ using FORCAST/SOFIA. Although the original AGN sample is larger than that presented here, only $22$ AGN were observed in total by SOFIA before end of operations in 2022. Note that there are gaps in the $20-214~\mu$m wavelength range due to the fact that SOFIA only flies with a single instrument per night. For each SOFIA cycle, we prioritized the objects with observations acquired in a single instrument from the previous cycle.

\begin{figure}
    \centering
    \includegraphics[width=\columnwidth]{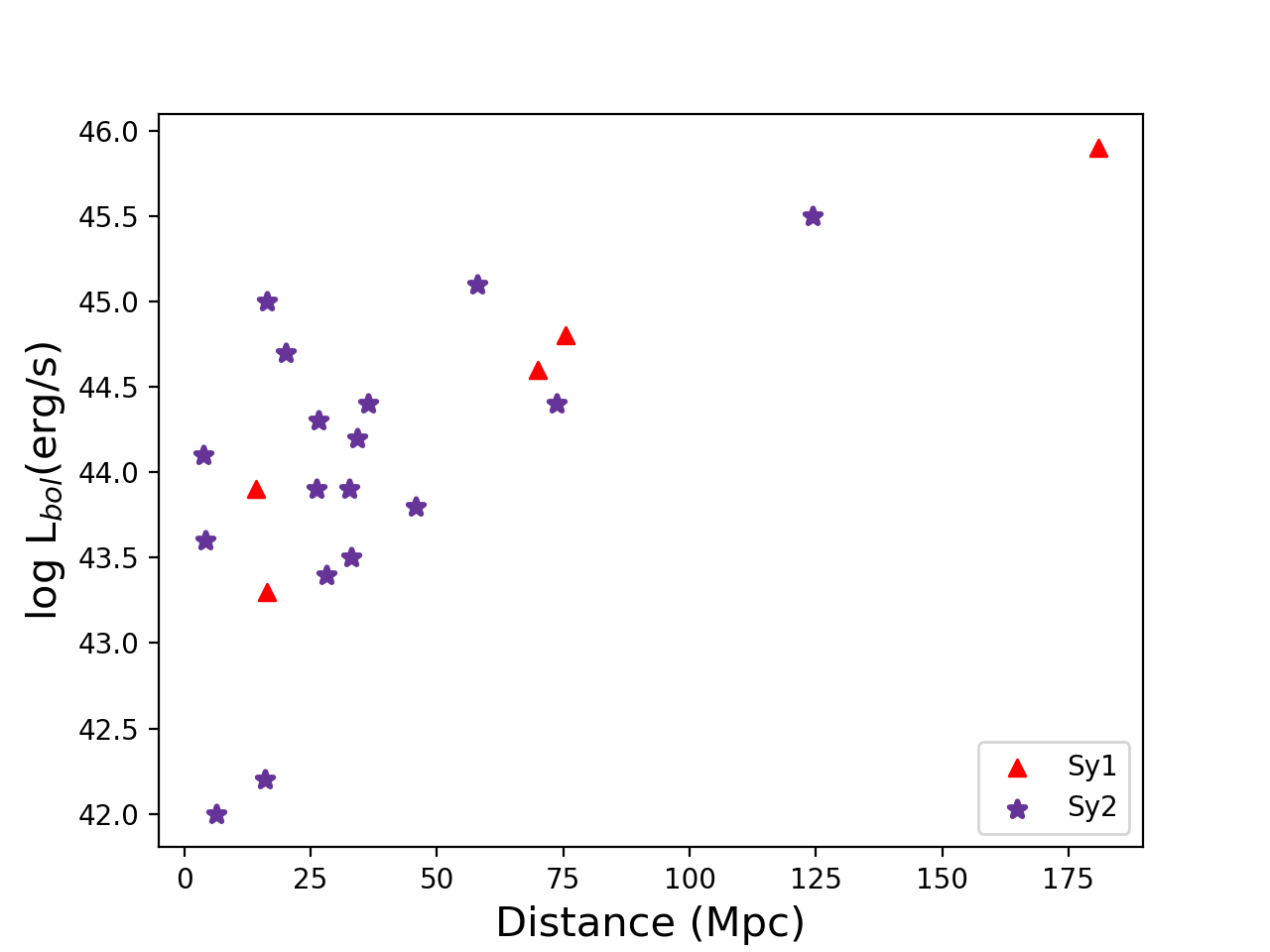}
    \caption{Luminosity plotted against distance of the 22 AGN in the atlas. Red triangles represent Sy1 and purple stars represent Sy2.}
    \label{fig:LvD}
\end{figure}

The sample properties are given in Table \ref{properties}. For most objects, we retrieved redshift data from the NASA Extragalactic Database (NED). However, for nearby objects Centaurus A and Circinus, distances were obtained individually \citep{Harris2010,Tully2009}. The 22 objects in this atlas cover the luminosity range of $\log_{10} (L_{\rm bol}[\rm{erg/s}]) = [42,46]$ with a median of $\log_{10} (L_{\rm bol}[\rm{erg/s}]) = 44.1\pm1.0$, and a distance of $4-181$ Mpc with a median of 42.8 Mpc.  Figure \ref{fig:LvD} shows bolometric luminosity plotted against distance, where Seyfert 1 objects are shown as red triangles and Seyfert 2 objects are shown as purple stars.

\begin{table}
\centering
\caption{SOFIA AGN Sample}
\scriptsize{
\label{properties}
\begin{tabular}{lcccccc}
\hline
Object	&	Type		&	$z$	&	Dist.  	&	Scale	&	log$L_{\rm{Bol}}$	&	Refs.	\\
		&			&		&	(Mpc)	&	 (pc/")	&	(erg s$^{-1}$)    &	 \\
\hline

CentaurusA	&	RLSy2	& 0.0018	&	3.8		&	17	&	 44.0		&	a\\
Circinus		&	Sy2		& 0.0014	&	4.2		&	19	&	43.6	&b	\\
MCG-5-23-16	&	Sy2		& 0.0085	&	36.4		&	177	&	44.4	&	c\\
Mrk 3		&	Sy2		& 0.0135	&	57.9		&	280	&	45.1	&	d\\
Mrk 231		&ULIRG/	& 0.0422 	&	181	&	877	&	45.9		&	e\\
            & Sy1   &           &           &       &               &    \\
Mrk 573		&	Sy2		& 0.0172	&	73.7		&	357	&	44.4	&b	\\
NGC 1068	&	Sy2		& 0.0038	& 	16.3		&	79	&	45.0	&f	\\
NGC 1275	&	RG/ 	& 0.0176	&	75.4		&	366	&	44.8 &g	\\
            &   Sy1.5   &           &               &       &       &   \\
NGC 2110	&	Sy2		& 0.0076	&	32.6		&	158	&	43.9	&b	\\
NGC 2273	&	Sy2		& 0.0061	&	26.1		&	127	&	43.9	&b	\\
NGC 2992	&  Sy1.9	& 0.0077	&	33.0		&	160	&	43.5&h	\\
NGC 3081	&	Sy2		& 0.0080	&	34.2 		&	166	&	44.2	&b	\\
NGC 3227	&	Sy1.5	& 0.0038	&	16.3 		&	79	&	43.3	&i	\\
NGC 3281	&	Sy2		& 0.0107	&	45.9		&	222	&	43.8	&b	\\
NGC 4151	&	Sy1.5	& 0.0033	&	14.1		&	69	&	43.9	&f	\\
NGC 4258    &   Sy2     & 0.0015   &   6.4         &   31  &   42.0    &j   \\
NGC 4388	&	Sy2		& 0.0047	&	20.1		&	98	&	44.7	&d	\\
NGC 4941	&	Sy2		& 0.0037	&	 15.9 	&	77	&	42.2	&k	\\
NGC 5506	&	Sy1.9	& 0.0062	&	26.6  	&	129	&	44.3	&b	\\
NGC 7465	&	Sy2/	& 0.0066	&	28.1		&	136	& 43.4 &k	\\
            &   LINER   &           &               &       &   &       \\
NGC 7469	&	Sy1		& 0.0163	&	69.9		&	339  &	44.6&i	\\
NGC 7674	&	Sy2		& 0.0290	&	124.3 	&	603	&	45.5	&b	\\
\hline 

\end{tabular}\\
}
Redshifts and spectral type were taken from NED. Distances to most sources were obtained using H$_{0}$ = 70 km s$^{-1}$ Mpc$^{-1}$.  Distances to nearby sources Centaurus A and Circinus were taken from \citet{Harris2010} and \citet{Tully2009}, respectively. References for log $L_{bol}$: a) \citet{Borkar2021} b) \citet{Marinucci2012} c) \citet{AH2011} d) \citet{Ichikawa2017}, e) \citet{Leighly2014} f) \citet{Marconi2004}, g) \citet{Baumgartner2013}, h) \citet{GB2015}, i) \citet{RA2011}, j) \citet{Yuan2002}, k) \citet{Duras2020},  
\end{table}

\subsection{SOFIA Observations and Data Reduction}
\label{observations}

\subsubsection{FORCAST}
\label{forcast}
FORCAST is an IR camera and spectrograph sensitive in the wavelength range $5 - 40~\mu$m with a field of view (FOV) of 3.4\arcmin$\times$3.2\arcmin~and pixel scale 0.768 \arcsec/pixel. With one exception (NGC 1068 in the 19.7 $\mu$m filter), we only used the Long Wavelength Camera (LWC; $25 - 40~\mu$m) due to the abundance of ground-based images at shorter wavelengths for the objects in our sample. FORCAST observations were made in dual channel mode using the two-position chop-nod (C2N) method with symmetric nod-match-chop (NMC) to remove telescope thermal emission and time variable sky background, and to reduce the effect of 1/$f$ noise from the array. Data were reduced by the SOFIA Science Center using the \textsc{forcast\_redux} pipeline following the methods described by \citet{Herter2012}. Most of the pipeline changes over the cycles were to refine the spectroscopic mode of FORCAST with little or no change to the image mode presented here.

Observations were flux-calibrated using the set of standard stars of the observing run, which provides flux uncertainties of $\sim$10 \%. The point spread function (PSF) of the 31.5 $\mu$m observations from Cycle 2 \cite[see][]{F16} was the co-average of a set of standard stars from that cycle. Its FWHM was 3.40\arcsec, in agreement with the SOFIA Observer's Handbook v3.0.0. The PSFs for Cycle 4 in the 30 - 40 $\mu$m wavelength range were determined by using standard star observations from the individual flights \cite[see][]{F19} and averaged at FWHM $\sim$ 4.33\arcsec~and 4.58\arcsec~in the 31.5 and 37.1 $\mu$m filters, respectively. The FORCAST PSFs for the observations of NGC1068 are detailed in \citet{ELR2018}.  

\subsubsection{HAWC+}
\label{hawc}
HAWC+ is a FIR imaging polarimeter designed to allow total and polarized intensity imaging observations in four broad bands centered at $53$, $89$, $155$, and $214$ $\mu$m, corresponding to Bands A, C, D, and E respectively (see Table \ref{filters}). On-the-fly mapping (OTFMAP) observing modes were used for both imaging polarimetry and total intensity imaging. Observing modes for the individual observations are given in Table \ref{tab:observations}. Data taken in polarization mode was reduced using the  \textsc{hawc\_dpr pipeline} and the reduction steps presented in \citet{ELR2022a}. The Comprehensive Reduction Utility for SHARC II \citep[CRUSH;][]{Kovacs2006,Kovacs2008} was used to obtain the total intensity observations. HAWC+ observations were reduced following the same reduction steps. We quote the pipeline versions or CRUSH versions in Table \ref{tab:observations} to differentiate between the observing modes. There are no differences between CRUSH versions to obtain the total intensity images as all  the changes in the pipeline were done for the polarimetric mode. Table \ref{tab:observations} also shows the pixel scale of each image.

As in FORCAST observations, the source of uncertainty in the photometry for HAWC+ results from calibration factors of the standard stars associated with the observation, giving an uncertainty of $\sim$10 \% \citep{ELR2022a}. HAWC+ PSFs were estimated using standard star observations in 2017. Pallas was observed in Bands A and C on 7 November 2017, while Neptune was observed in Bands D and E on 19 October, 2017. The FWHM of these standards are 5.25\arcsec, 8.26\arcsec, 14.74\arcsec, and 19.65\arcsec~in Bands A, C, D, and E, respectively. The FWHMs from the Observer's Handbook\footnote{https://irsa.ipac.caltech.edu/data/SOFIA/docs/sites/default/files/Other/Documents/OH-Cycle7.pdf} are given in Table \ref{filters}. 

\begin{table}
\caption{HAWC+ filter suite}
\label{filters}
\begin{tabular}{lccc}

\hline
Filter  &   $\lambda_{central}$ ($\mu$m)  & $\Delta\lambda$ ($\mu$m) & FWHM (")  \\
\hline
Band A  &   53  &   8.7 &   4.85    \\
Band C  &   89  &   17  &   7.8     \\
Band D  &   154 &   34  &   13.6    \\
Band E  &   214 &   44  &   18.2    \\
\hline

\end{tabular}
\end{table}

\subsection{Observing Data}
\label{sec:data}
Table \ref{tab:observations} provides the final AGN sample with information about the wavelength, observing mode, observation and mission details, versions of the separate pipelines, and also the field-of-view (FOV) of the individual images.

%
\begin{longrotatetable}
\centering

\begin{deluxetable*}{ccccccccccccc}



\startdata
Object	&	Wavelength  &   Instrument & Date  & Observing mode    & Pipeline/ &	On-source 	&	Altitude	&	Mission ID		&	Program ID	&	Image FOV	& Pixel scale\\
		&	($\mu$m)	&           	& YYYY-MM-DD &	&  CRUSH vers. & time (s)		&	(ft)		&				&				&	(arcsec$^{2}$)	& (") \\
\hline
Centaurus A	&	53	&   HAWC+    &	2019-07-17	&polarization   &  2.3.2 & 	428	&	39022	&	HA\_F597	&	07\_0032	&	75"$\times$75"	& 1.29	\\
			&	89	&   HAWC+    &	2019-07-17	& polarization & 2.3.2 &	856	&	39020	&	HA\_F597	&	07\_0032	&	130"$\times$130" & 2.02	\\
\hline
Circinus	&	53	&   HAWC+    &	2019-07-16	&  polarization & 2.3.2	& 107	&	42014	&	HA\_F596	&	07\_0032	&	70"$\times$70"	 &1.00	\\
			&	89	&   HAWC+    &	2019-07-16	&  polarization & 2.3.2 &	856	&	40994	&	HA\_F596	&	07\_0032	&	105"$\times$105" & 2.00	\\
			&	215	&   HAWC+    &	2019-07-16	& polarization & 2.3.2 &	281	&	43011	&	HA\_F596	&	07\_0032	&	140"$\times$140" & 4.72	\\
\hline
MCG-5-23-16	&	31.5	& FORCAST    &	2014-05-03	&  imaging      & 1.0.1Beta	&324	&	38076	&	FO\_F167	&	02\_0035	&	20"$\times$20"	& 0.77	\\
			&	53	&   HAWC+     &	2018-07-04	& tot. intensity &  CSH 2.41-3	&321	&	39996	&	HA\_F480	&	06\_0066	&	50"$\times$50"		& 1.00 \\
			&	89	&   HAWC+     &	2018-07-04	& tot. intensity &  CSH 2.41-3   &	321	&	40003	&	HA\_F480	&	06\_0066	&	75"$\times$75"	&	1.55\\
			&	155	&   HAWC+     &	2018-07-04	& tot. intensity &  CSH 2.41-3	& 642	&	39991	&	HA\_F480	&	06\_0066	&	90"$\times$90"	& 2.75\\
\hline   
Mrk 3		&	31.5 &  FORCAST      &	2016-09-27	&  imaging       & 1.1.3 &	35	&	42994	&	FO\_F133	&	04\_0048	&	20"$\times$20"	& 0.77	\\
			&	37.1 &  FORCAST  	&	2016-09-27	& imaging       & 1.1.3 &	43	&	43002	&	FO\_F133	&	04\_0048	&	20"$\times$20"	& 0.77	\\
			&	53	&   HAWC+     &	2018-09-21	& tot. intensity & CSH 2.41-3    &	1926	&	43002	&	HA\_F508	&	06\_0066	&	50"$\times$50"	& 1.00	\\
			&	89	&   HAWC+     &	2019-02-13	& tot. intensity & CSH 2.41-3    &	1277	&	43015	&	HA\_F546	&	06\_0066	&	75"$\times$75"	&  1.55	\\
			&	155	&   HAWC+     &	2019-02-13	& tot. intensity & CSH 2.41-3    &	766	&	43003	&	HA\_F546	&	06\_0066	&	90"$\times$90"		& 2.75\\
\hline
Mrk 231		&	89	&   HAWC+    &	2019-09-18	& polarization  & 2.3.2 &	749	&	40010	&	HA\_F611	&	07\_0032	&	75"$\times$75"	& 2.00	\\
\hline
Mrk 573		&	31.5 &  FOR  	&	2015-02-05	& imaging       & 1.0.3 &	384	&	40008	&	FO\_F192	&	02\_0035	&	20"$\times$20"	& 0.77	\\
			&	37.1 &  FOR  	&	2018-09-08	& imaging       & 1.3.2 &	31	&	43006	&	FO\_F502	&	06\_0066	&	20"$\times$20"	& 0.77	\\
   \hline
NGC 1068	&	19.7 &  HAWC+ 	&	2016-09-17	& imaging        & 1.1.3 &	427	&	42980	& FO\_F329	&	70\_0400	&	30"$\times$30"	& 0.77	\\
			&	31.5 &  HAWC+ 	&	2016-09-17	& imaging       & 1.1.3 &	471	&	42980	&	FO\_F329	&	70\_0400	&	40"$\times$40"	& 0.77\\
			&	37.1 &  HAWC+ 	&	2016-09-17	&	imaging       & 1.1.3 &343	&	42980	&	FO\_F329	&	70\_0400	&	40"$\times$40" &	0.77	\\
			&	53	&   HAWC+     &	2016-12-08	& tot. intensity & CSH 2.41-3 &	455	&	40012	&	HA\_F356	&	70\_0409	&	60"$\times$60"	& 1.00	\\
			& 89    &   HAWC+    &   2017-10-20  & polarization & 2.7.0 &  1364    & 40018     &   HA\_F443 &   08\_0012     &  10"$\times$60" & 4.02\\
\hline   
NGC 1275	&	31.5 &  FORCAST      &	2016-09-21	& imaging & 1.1.3 &	48	&	41103	&	FO\_F274	&	04\_0048	&	20"$\times$20"	& 0.77	\\
			&	37.1 &  FORCAST      &	2016-09-21	& imaging & 1.1.3 &	56	&	41094	&	FO\_F274	&	04\_0048	&	20"$\times$20"	& 0.77\\
            & 89    & HAWC+ &   2019-09-05 & polarization & 2.3.2 & 7680 & 40021 & HA\_F606 & 07\_0032 & 50"$\times$50"  &   1.55 \\
\hline   
NGC 2110	&	31.5 &  FORCAST      &	2015-02-05	& imgaging & 1.0.3 &	768	&	38001	&	FO\_F192	&	02\_0035	&	20"$\times$20" &	0.77	\\
			&	53	 &  HAWC+    &	2018-09-28	& tot. intensity & CSH 2.41-3 &	642	&	42984	&	HA\_F512	&	06\_0066	&	50"$\times$50"	& 1.00	\\
			&	89	&   HAWC+     &	2018-09-28	& tot. intensity & CSH 2.41-3 & 320	&	42945	&	HA\_F512	&	06\_0066	&	75"$\times$75"	& 1.55	\\
			&	155	&   HAWC+     &	2018-09-28	& tot. intensity & CSH 2.41-3 &	320	&	42955	&	HA\_F512	&	06\_0066	&	90"$\times$90"	& 2.75	\\
			&	215	&   HAWC+     &	2018-09-28	& tot. intensity & CSH 2.41-3 &	320	&	42945	&	HA\_F512	&	06\_0066	&	90"$\times$90"	&3.70	\\
\hline   
NGC 2273	&	31.5 &  FORCAST      &	2016-09-27	& imaging & 1.1.3 &	34	&	42998	&	FO\_F333	&	04\_0048	&	20"$\times$20"	& 0.77 \\
			&	37.1 &  FORCAST      &	2016-09-27	& imaging & 1.1.3 &	41	&	42996	&	FO\_F333	&	04\_0048	&	20"$\times$20"	& 0.77 \\
			&	53	&   HAWC+     &	2018-09-28	& tot. intensity & CSH 2.41-3 &	428	&	38987	&	HA\_F512	&	06\_0066	&	50"$\times$50"	& 1.00\\
			&	89	&   HAWC+     &	2018-09-28	& tot. intensity & CSH 2.41-3 &	534	&	38949	&	HA\_F512	&	06\_0066	&	75"$\times$75"	& 1.55\\
			&	155	&   HAWC+     &	2018-09-28	& tot. intensity & CSH 2.41-3 &	321	&	38956	&	HA\_F512	&	06\_0066	&	90"$\times$90"	& 2.75\\
			&	215	&   HAWC+     &	2018-09-28	& tot. intensity & CSH 2.41-3 &	321	&	38960	&	HA\_F512	&	06\_0066	&	90"$\times$90"	& 3.70\\
 \hline  
NGC 2992	&	31.5 &  FORCAST      &	2014-05-02	& imaging & 1.0.1Beta &	232	&	39026	&	FO\_F166	&	02\_0035	&	20"$\times$20"	& 0.77	\\
\hline
NGC 3081	&	31.5 &  FORCAST      &	2014-05-02	& imaging & 1.0.1Beta &	480	&	39021	&	FO\_F166	&	02\_0035	&	20"$\times$20"	& 0.77	\\
			&	37.1 &  FORCAST      &	2016-02-17	& imaging & 1.0.1Beta &	63	&	43004	&	FO\_F278	&	04\_0048	&	20"$\times$20"	& 0.77\\
 \hline  
NGC 3227	&	31.5 &  FORCAST      &	2014-05-06	& imaging & 1.0.1Beta &	152	&	37990	&	FO\_F168	&	02\_0035	&	20"$\times$20"	& 0.77	\\	
			&	37.1 &  FORCAST      &	2016-02-17	& imaging & 1.1.0 &	35	&	43002	&	FO\_F278	&	04\_0048	&	20"$\times$20"	& 0.77	\\
\hline   
NGC 3281	&	31.5 &  FORCAST      &	2014-05-02	& imaging & 1.0.1Beta &	300	&	39033	&	FO\_F166	&	02\_0035	&	20"$\times$20"	& 0.77\\
			&	53	 &  HAWC+     &	2018-07-12	& tot. intensity & CSH 2.41-3 &	321	&	37036	&	HA\_F485	&	06\_0066	&	50"$\times$50"	& 1.00	\\
			&	89	&   HAWC+     &	2018-07-12	& tot. intensity & CSH 2.41-3 &	320	&	37025	&	HA\_F485	&	06\_0066	&	75"$\times$75"	& 1.55\\
			&	155	&   HAWC+     &	2018-07-12	& tot. intensity & CSH 2.41-3 &	320	&	37043	&	HA\_F485	&	06\_0066	&	90"$\times$90"	& 2.75\\
			&	215	&   HAWC+     &	2018-07-12	& tot. intensity & CSH 2.41-3 &	534	&	37049	&	HA\_F485	&	06\_0066	&	90"$\times$90"	& 3.70\\
\hline   
NGC 4151	&	31.5 &  FORCAST      &	2014-05-08	& imaging & 1.0.1Beta & 153	&	40042	&	FO\_F170	&	02\_0035	&	20"$\times$20"	& 0.77\\
			&	37.1 &  FORCAST      &	2016-02-17	& imaging & 1.0.1Beta &	41	&	36996	&	FO\_F278	&	04\_0048	&	20"$\times$20"	& 0.77	\\
			&	53	 &  HAWC+     &	2019-02-12	& tot. intensity & CSH 2.41-3 &	1069	&	39777	&	HA\_F545	&	06\_0066	&	50"$\times$50"	&	1.00\\
			&	89	 &  HAWC+     &	2019-02-12	& polarization & 2.3.2 &	891	&	40034	&	HA\_F545	&	06\_0066	&	75"$\times$75"	& 1.55\\
\hline   
NGC 4258    &   31.5 &  FORCAST      &  2016-02-17  & imaging & 1.1.0 &   156 &   39002   &   FO\_F278    &   04\_0048    &   30"$\times$30"  & 0.77 \\
            &   37.1 &  FORCAST      &  2016-02-17  &  imaging & 1.1.0 &  446 &   38999   &   FO\_F278    &   04\_0048    &   30"$\times$30"  & 0.77 \\
\hline            
NGC 4388	&	31.5 &  FORCAST      &	2014-05-02	& imaging & 1.0.1Beta &	162	&	37982	&	HA\_F545	&	02\_0035	&	20"$\times$20"	& 0.77 \\
			&	37.1 &  FORCAST      &	2016-02-06	& imaging & 1.1.0 &	63	&	42372	&	FO\_F274	&	04\_0048   &	20"$\times$20" &	0.77\\
			&	53	 &  HAWC+     &	2019-02-12	& tot. intensity & CSH 2.41-3 &	932	&	42045	&	HA\_F545	&	06\_0066	&	50"$\times$50"	& 1.00\\
			&	89	 &  HAWC+     &	2019-02-12	& tot. intensity & CSH 2.41-3 &	534	&	42035	&	HA\_F545	&	06\_0066	&	75"$\times$75"	& 1.55\\
			&	155	 &  HAWC+     &	2019-02-12	& tot. intensity & CSH 2.41-3 &	321	&	42021	&	HA\_F545	&	06\_0066	&	90"$\times$90"	& 2.00	\\
			&	215	 &  HAWC+     &	2019-02-12	& tot. intensity & CSH 2.41-3 &	213	&	42037	&	HA\_F545	&	06\_0066	&	90"$\times$90"	& 3.70 \\
\hline   
NGC 4941	&	31.5 &  FORCAST      &	2021-04-09	& imaging & 2.2.1 &	154	&	38992	&	FO\_F715	&	08\_0014	&	20"$\times$20" & 0.77 \\
		&	37.1 &  FORCAST      &	2021-04-09	& imaging & 2.2.1 & 313	&	38986	&	FO\_F715	&	08\_0014	&	20"$\times$20"	& 0.77\\
 \hline 
NGC 5506	&	31.5 &  FORCAST      &	2014-05-02	& imaging & 1.0.1Beta &	261	&	37982	&	FO\_F166	&	02\_0035	&	20"$\times$20"	& 0.77	\\
		&	37.1 &  FORCAST      &	2021-06-30	& imaging & 2.3.0 &	427	&	39005	&	FO\_F752	&	08\_0014	&	20"$\times$20"	& 0.77\\
        &   89  & HAWC+ &   2020-01-31  &   polarization & 2.3.2 & 779  &   43009   &   HA\_F657    & 07\_0032 & 75"$\times$75" & 1.55 \\
  \hline
NGC 7465    &   31.5    &   FORCAST &   2022-09-17  & imaging & 2.6.0 &  427 &   41017   &   FO\_F915 &  08\_0014    &   20"$\times$20"  & 0.77\\
            &   37.1    & FORCAST &  2022-09-17 & imaging & 2.6.0 &  222 &   41009      &    FO\_F915    & 08\_0014    &   20"$\times$20" & 0.77 \\
            &   53   &  HAWC+     &   2021-09-14  & tot. intensity & CSH 2.42-1 &  641 &   38008   &   HA\_F783    &   08\_0014    &   20"$\times$20"  & 1.00\\
            &   89   &  HAWC+     &   2021-09-14  & tot. intensity & CSH 2.42-1 &   426 &   39004   &   HA\_F783    &   08\_0014    &   20"$\times$20"  & 1.55 \\
\hline
NGC 7469	&	31.5 &  FORCAST      &	2014-06-04	& imaging & 1.0.1Beta &	231	&	43005	&	FO\_F176	&	02\_0035	&	20"$\times$20"	& 0.77	\\
\hline
NGC 7674	&	31.5 &  FORCAST      &	2014-06-04	& imaging & 1.0.1Beta &	165	&	43005	&	FO\_F176	&	02\_0035	&	20"$\times$20"	& 0.77	\\
\enddata
\tablecomments{Observation Data - Column 1: Object; Column 2: Wavelength; Column 3: Instrument; Column 4: Observation date; Column 5: Observing mode; Column 6: Pipeline/CRUSH vershion; Column 7: On-source time; Column 8: Aircraft starting altitude; Column 9: Mission ID; Column 10: Program ID; Column 11: FOV of images in Section \ref{sec:images}; Column 12: Pixel scale}
\label{tab:observations}
\end{deluxetable*}
\end{longrotatetable}


\section{Images}
\label{sec:images}
 
\begin{figure*}
\centering
\includegraphics[scale=0.7]{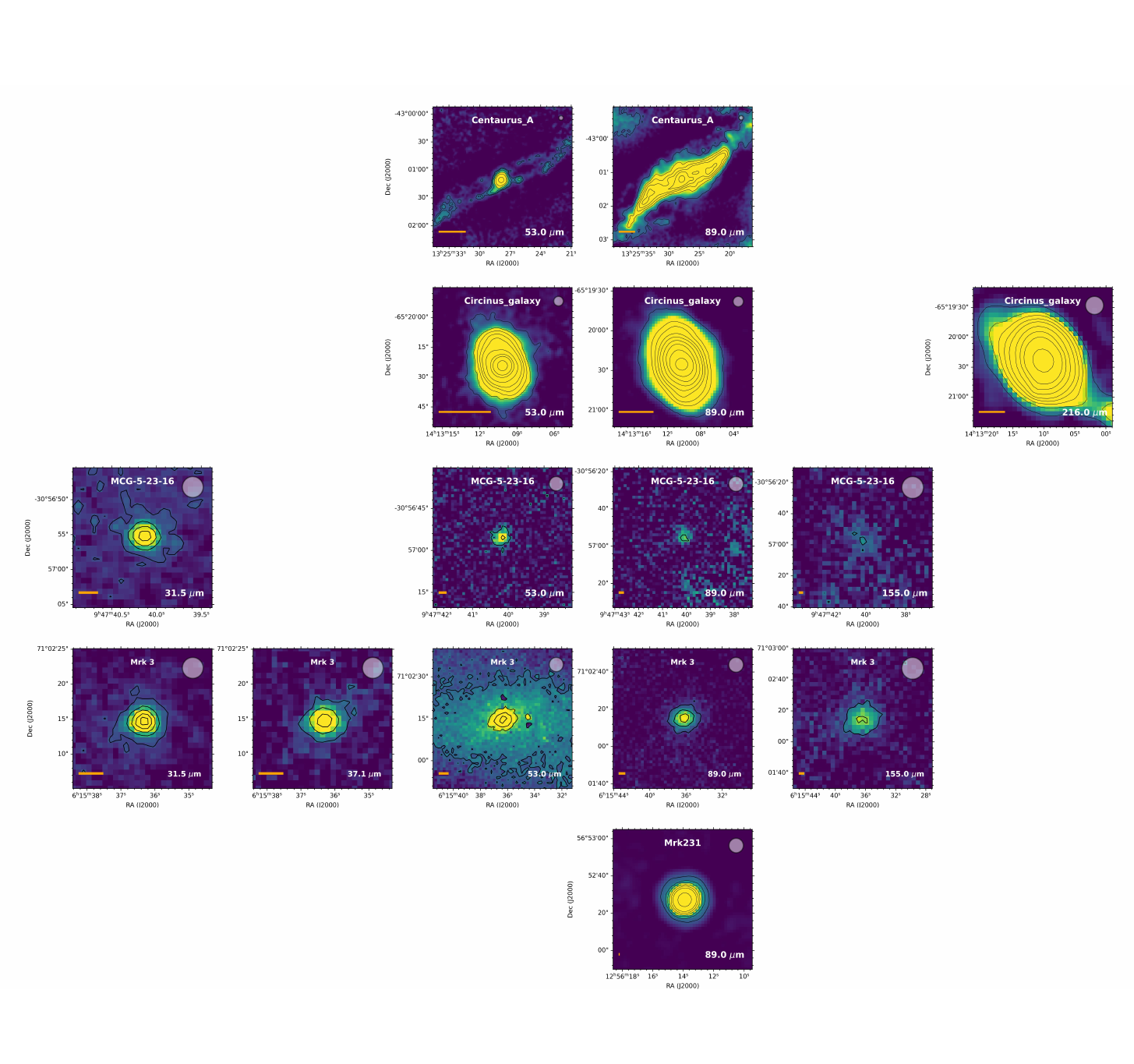}
\caption{FORCAST 31.5, and 37.1 $\mu$m images, and HAWC+ 53, 89, 155, and 215 $\mu$m images.  Each image has a differing FOV, which can be found in Table \ref{tab:observations}. For bright objects (Centaurus A, Circinus, and Mrk 231) contours start at 3$\sigma$, then follow log(maximum) from [-1.2 to 0.8], [-1.8 to 0.8] [-1.4,0.8] in steps of 0.2. For MCG-5-23-16 and Mrk 3, the lowest contours are 3$\sigma$ and increase in steps of 5$\sigma$. The white transparent circle on the top right indicates the telescope beam size. The orange bar on the bottom left of the images is scaled to 500 pc.  For all images, north is up and east is to the left.}
\label{image1}
\end{figure*}

\begin{figure*}
\centering
\includegraphics[scale=0.7]{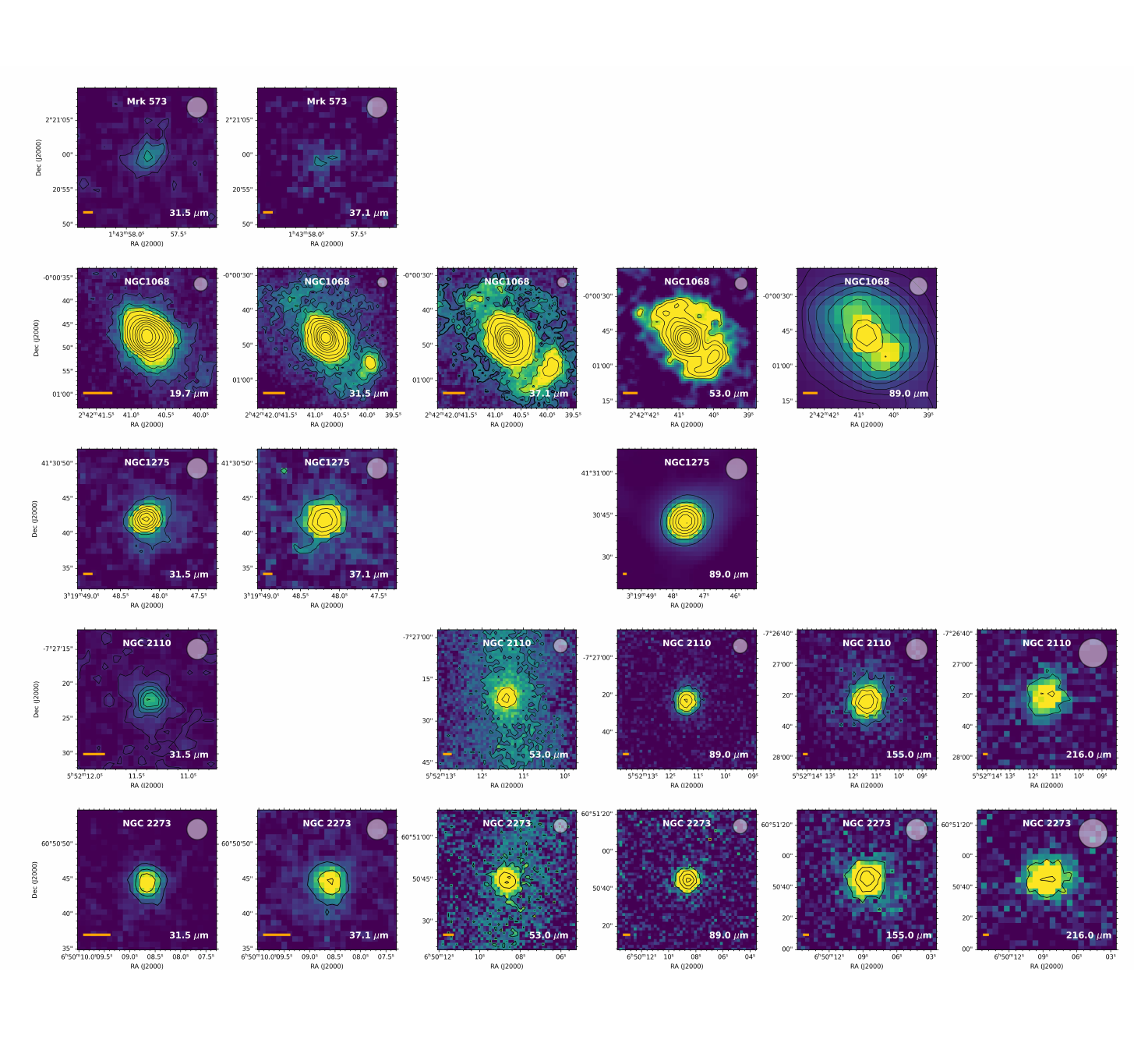}
\caption{FORCAST 31.5 and 37.1 $\mu$m images, and HAWC+ 53, 89, 155, and 215 $\mu$m images. Note that the wavelength range for NGC 1068 starts 19.7 $\mu$m, so its range is shifted.  Each image has a differing FOV, which can be found in Table \ref{tab:observations}. For NGC 1068 FORCAST images, contours begin at 3 $\sigma$ and follow log(maximum) from [-2.0,0.8] in steps of 0.2, while the HAWC+ images follow the same steps but log (max) ranges [-1.6,0.8].  For all other images, the lowest contours are 3$\sigma$ and increase in steps of 5$\sigma$. The white transparent circle on the top right indicates the telescope beam size. The orange bar on the bottom left of the images is scaled to 500 pc. For all images, north is up and east is to the left.}
\label{image2}
\end{figure*}

\begin{figure*}

\includegraphics[scale=0.7]{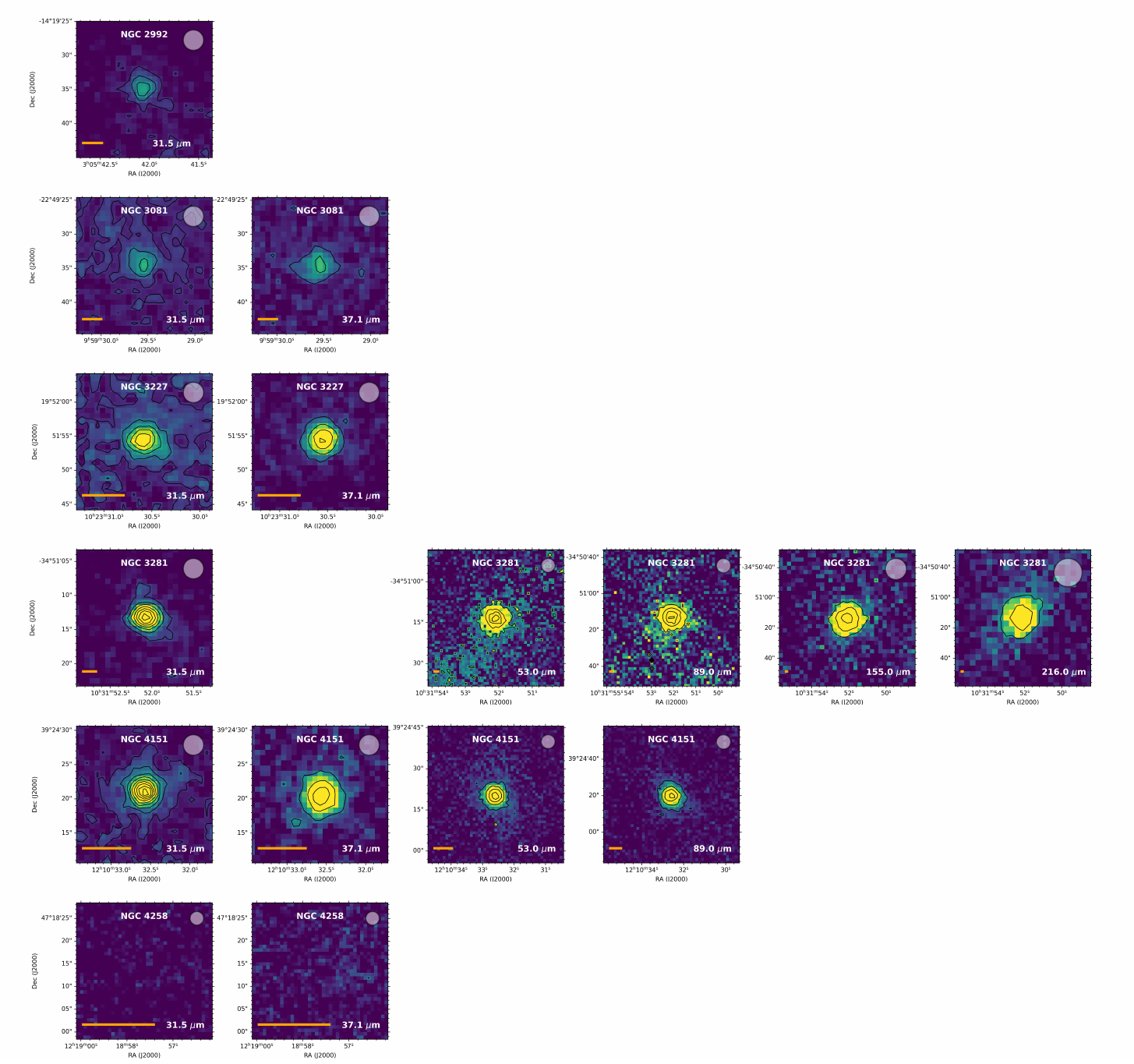}
\caption{FORCAST 31.5 and 37.1 $\mu$m images, and HAWC+ 53, 89, 155, and 215 $\mu$m images.  Each image has a differing FOV, which can be found in Table \ref{tab:observations}. The lowest contours are 3$\sigma$ and increase in steps of 5$\sigma$. The white transparent circle on the top right indicates the telescope beam size. The orange bar on the bottom left of the images is scaled to 500 pc. For all images, north is up and east is to the left. }
\label{image3}
\end{figure*}

\begin{figure*}

\includegraphics[scale=0.7]{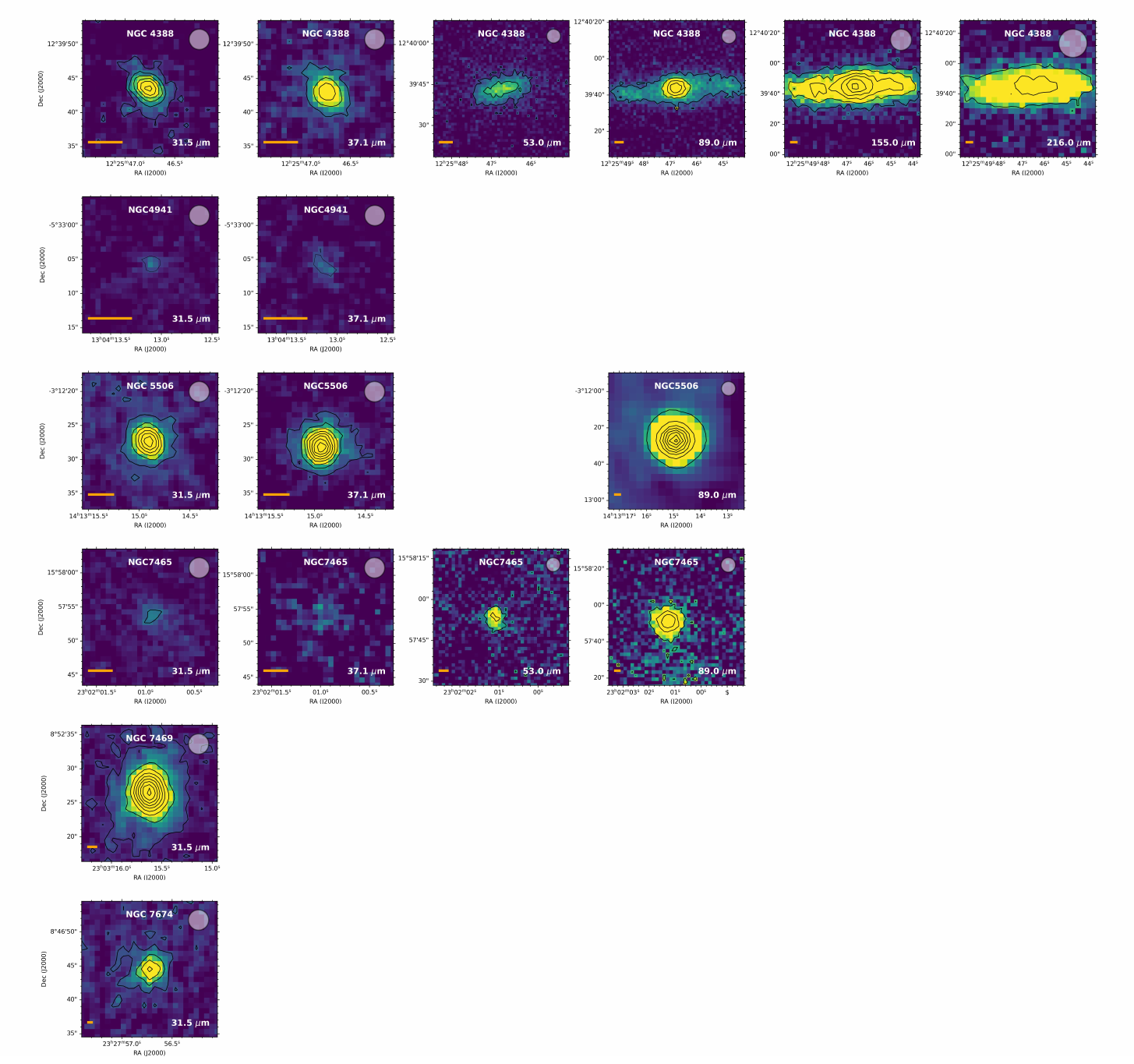}
\caption{FORCAST 31.5 and 37.1 $\mu$m images, and HAWC+ 53, 89, 155, and 215 $\mu$m images.  Each image has a differing FOV, which can be found in Table \ref{tab:observations}. The white transparent circle on the top right indicates the telescope beam size. The orange bar on the bottom left of the images is scaled to 500 pc. For all images, north is up and east is to the left. }
\label{image4}
\end{figure*}

Images of the $22$ AGN in the $19.7 - 214~\mu$m wavelength range are presented in Figures \ref{image1}, \ref{image2}, \ref{image3}, and \ref{image4}. The orange scale on the bottom left of the images indicates a scale of $500$ pc. The beam size is depicted in white in the top right of the images. In all images, north is up and east is to the left. Complementary \textit{Herschel} $70 - 500~\mu$m images are shown in Appendix \ref{herschel_images} (Fig. \ref{appendix:herschel1}, \ref{appendix:herschel2}, \ref{appendix:herschel3}, \ref{appendix:herschel4}). These fully reduced images were obtained through the \textit{Herschel} Science Archive\footnote{The \textit{Herschel} archive can be found at \url{http://herschel.esac.esa.int/Science_Archive.shtml}}. All objects are presented and analyzed individually.

\textbf{Centaurus A}.
The host galaxy of Centaurus A is clearly visible in the 53 $\mu$m image with a bright compact center. However, the host galaxy becomes more dominant in the 89 $\mu$m image \citep[see a detailed analysis of host galaxy dust emission at 89 $\mu$m in][]{ELR2021}. The kpc-scale warped dust and gas lane was first observed with \textit{Spitzer} imaging using IRAC and MIPS \citep{Quillen2006}. On subarcsecond scales, \citet{Radomski2008} observed the nucleus of Centaurus A using the 8.8, 10.4, and 18.3 $\mu$m filters on T-ReCS at Gemini South. They concluded that the mostly likely sources of nuclear MIR emission are an unresolved clumpy dusty torus in the core, and a dusty NLR for the arcsecond-scale extended emission \citep[see also][]{GBernete2016}. 
\\

\textbf{Circinus Galaxy}.
HAWC+ 53 and 89 $\mu$m images of the Circinus Galaxy show a very bright FIR core with extended emission at a PA $\sim$ 30$^{\circ}$, whereas the 215 $\mu$m image shows a slightly different PA $\sim$ 55$^{\circ}$. We estimate the FWHM of the extended FIR nuclear emission to be $\sim$ 6\arcsec$\times$6\arcsec, 13.5\arcsec$\times$11.5\arcsec, and 22.3\arcsec$\times$25.6\arcsec~at 53, 89, and 214 $\mu$m, respectively. These are larger than the PSF FWHMs given in Section \ref{hawc}, which indicates extended emission along the axis of the inner bar of the galaxy. MIR emission was resolved at 8.7 and 18.3 $\mu$m out to $2\arcsec$ in an approximate east-west direction, coincident with the ionization cones at PA $\sim$ 100$^{\circ}$ \citep{packham2005,Stalevski2017}. However, the elongation seen in the SOFIA images (and in \textit{Herschel} images in Appendix \ref{herschel_images}) seems to be arising from dust in the host galaxy. 

\textbf{MCG-5-23-16}.
MCG-5-23-16 appears as a point-like source in the $31.5-155~\mu$m wavelength range whose brightness decreases with increasing wavelength. However, this galaxy appears as an extended source in the MIR using high angular resolution data from VLT \citep{GBernete2016}.  Likewise, \citet{Ferruit2000} found that this nucleus has an extended optical NLR at a PA of $40^{\circ}$. They  found a dust lane extending $2\arcsec$ on either side of the nucleus, parallel to the axis of the galaxy. Their extended dusty emission at $40^{\circ}$ was detected at a $3\sigma$ level up to  $\sim4\arcsec$ from the core. This extended structure has no thermal emission counterpart within the $31.5-155$ $\mu$m wavelength range.

\textbf{Mrk 3}.
Although the FORCAST and HAWC+ images generally appear to be point-like sources, \citet{F19} found extended emission in the PSF-subtracted 37.1 $\mu$m image of Mrk 3 in the direction of the radio axis \citep[84$^{\circ}$;][]{Kukula1993} and the NLR \citep[$\sim$ 70$^{\circ}$;][]{Capetti1995}. \citet{sasmirala} found an elongated nucleus out to $\sim$ 170 pc with PA $\sim$ 70$^{\circ}$ in the Si-5 (11.6 $\mu$m) filter using Gemini/MICHELLE. However, the Si-2 (8.7 $\mu$m) image from GTC/Canaricam appears point-like \citep{AH2016}. The large scale east-west structure in the 53 $\mu$m image here is a background artifact produced by the data reduction due to the small spatial coverage and short integration time of the observation. 

\textbf{Mrk 231}.
The 89 $\mu$m image shown here is point-like with a FWHM of $\sim$ 8.5", similar to the standard FWHM of 8.26" (see Section \ref{hawc}). Mrk 231 is a Type 1 Ultra Luminous Infrared Galaxy (ULIRG) and is the nearest known quasar at a distance of 181 Mpc. It is known for its multi-phase and multi-scale outflows \citep[see][]{Rupke2011}, with a neutral outflow up to 3 kpc in radius \citep{Rupke2005}. 
\textbf{Mrk 573}. 
Although the SNR is very low ($3-4\sigma$), both 31.5 and 37.1 $\mu$m images of Mrk 573 show marginally resolved $\sim4.5\arcsec$ elongation in the east-west direction at PA $\sim$ 110$^{\circ}$. Mrk 573 was previously shown to have a biconical NLR coincident with radio emission 3-4" from the nucleus at a PA $\sim$ 125$^{\circ}$ \citep{Ulvestad1984,Pogge1995}. The marginal detection here may be cold extended dust in the outer layers of the NLR. 

\textbf{NGC 1068}. 
The 19 - 53 $\mu$m images of NGC 1068 were published previously in \citet{ELR2018}, where it was shown that the peak emission from the torus occurs between 30 - 40 $\mu$m with a corresponding temperature of 70 - 100 K. The 89 $\mu$m image was published as a polarimetric observation \citep{ELR2020}. Within a scale of about $1$ kpc, NGC 1068 shows extended emission in the NE to SW direction at a PA$ \sim 45^{\circ}$, similar to MIR observations using VISIR/VLT \citep{Asmus2014}. Their observations revealed a nuclear structure in the north-south direction and extended structures to the NE and SW. From the N-band spectrum, \citet{Mason2006} concluded that while torus emission dominates NIR wavelengths, large-scale MIR emission is dominated by diffuse dust within the ionization cones.  

\textbf{NGC 1275}.
All 30 - 53 $\mu$m images are dominated by a point-like source. However, the 31.5 $\mu$m image \citep{F19} shows $3\sigma$ extended emission along the PA $\sim140^{\circ}$. This AGN is known to have a network of H$\alpha$ filaments extending out to $\sim$ 100" \citep[see][]{Conselice2001} and is possibly the result of a merger \citep{Holtzman1992}. The MIR core shows silicate dust emission in both 10 and 18 $\mu$m bands \citep[see][]{F19}. Hence, both dust and gas are extended covering several kpc around the core. In the HAWC+ 89 $\mu$m filter, \citet{ELR2023} found extended dust emission at a PA $\sim$ 125$^{\circ}$ out to a 12 kpc radius potentially associated with a magnetized dusty filament along the NW direction \citep{Fabian2008}.

\textbf{NGC 2110}. 
The 30 - 215 $\mu$m images of NGC 2110 are all point-like. The north-south pattern in the 53 $\mu$m image is due to background noise and does not represent extended dust. NGC 2110 is a Type 2 AGN that shows silicate emission at 10 and 18 $\mu$m that is interpreted as a result of a clumpy torus, or as dust within the ionization cones \citep[PA $\sim$ 160$^{\circ}$;][]{Mulchaey1994} in the inner 32 pc of the AGN \citep{Mason2009}. This galaxy appears as an extended source in Gemini/MICHELLE high resolution N-band observations \citep{GBernete2016}. However, any structure within the NLR or ionization cones is not resolved by our observations.

\textbf{NGC 2273}. 
The full set of 30 - 215 $\mu$m images of NGC 2273 show a point-like source. The north/south pattern in the 53 $\mu$m image is due to background noise and does not represent extended dust.  Within the FWHM of these images (see Sections \ref{forcast}, \ref{hawc}), there is a known star-forming ring within $\sim$ 2" of the nucleus \citep{Ferruit2000,Martini2003,Sani2012}. GTC/Canaricam observations \citep{AH2014,AH2016} at 8.7 $\mu$m show elongation from the north-east to the south-west, likely with contribution from PAH. This structure is consistent with extension seen in the PSF-subtracted 37.1 $\mu$m SOFIA image \citep{F19}. 

\textbf{NGC 2992}.
The image of NGC 2992 in the 31.5 $\mu$m filter is published in \citet{F16} and appears as a point-like source. Subarcsecond N-band imaging \citep{GB2015} reveals extended emission along PA $\sim$ 30$^{\circ}$ out to $\sim$ 3 kpc which is attributed to dust heated by star formation based on corresponding N-band spectroscopy.  The FWHM of the SOFIA image is $\sim$ 3.5"$\times$3.5" (560$\times$560 pc$^{2}$) so the extension should be resolvable within the SOFIA image. Since we do not see the extension in the image here, we conclude that either the extended dust emission tapers at wavelengths > 20 $\mu$m or SOFIA does not have enough sensitivity to detect it.

\textbf{NGC 3081}.
The 31.5 $\mu$m image of NGC 3081 was published in \citet{F16} while the 37.1 $\mu$m image was published in \citet{F19}. The nucleus is known to harbor a region of strong optical emission $\sim$ 1" from the AGN \citep{Ferruit2000} likely due to dust or gas heated by the AGN. \citet{F19} estimated that $\sim35$\% of the MIR emission within the central few arcseconds (few hundred parsecs) of the AGN originates in the NLR. High angular resolution N- and Q- band observations show extension towards the north, extending out to $\sim$450 pc from the south-east to the north-west \citep[PA$\sim$160$^{\circ}$;][]{GBernete2016}. On larger scales, optical and NIR observations reveal a series of star forming resonance rings at distances of 2.3, 11.0, 26.9 kpc and 33.1 kpc \citep{Buta1990,Buta1998,Buta2004}. At longer wavelengths (>200 $\mu$m), \citet{RA2011} concluded that FIR emission is contaminated by the star-forming ring 2.3 kpc in diameter.

\textbf{NGC 3227}.
The 31.5 $\mu$m image was published in \citet{F16} while the 37.1 $\mu$m image was published in \citet{F19}. These images show a point-like source, although NGC 3227 is known to harbor a nuclear star-forming region \citep{Schinnerer2001,Davies2006} with a nuclear cluster within $\sim$ 70 pc ($\sim$ 1") from the core. The 8.7 $\mu$m image from \citet{AH2016} shows a slight north/south elongation and the corresponding spectrum shows clear PAH in the nucleus \citep[see also][]{GBernete2016}.  These star forming regions likely contaminate the nuclear MIR emission within the FWHM of our images. 

\textbf{NGC 3281}.
While the 31.5 $\mu$m FORCAST image is published in \citet{F16}, the HAWC+ images at 53, 89, 154, and 214 $\mu$m are presented here for the first time and appear point-like in all filters. The images taken at 53 and 89 $\mu$m appear to have significant noise in their backgrounds. The subarcsecond (0.35") N-band spectrum in \citet{GM2013} shows a deep 10 $\mu$m silicate absorption feature which originates in the inner $\sim$ 80 pc of the AGN. 

\textbf{NGC 4151}.
The SOFIA images of NGC 4151 appear as a point-like source, with a potential detection of extended emission at PA$\sim120^{\circ}$ at a $3\sigma$ level at $37.1~\mu$m.  \citet{F19} confirmed this elongation in the PSF-subtracted 37.1 $\mu$m image coincident with the NLR and radio axes. \citet{Radomski2003} show extended emission in 10.8 and 18.2 $\mu$m images that coincides with the NLR axis at PA $\sim$ -60$^{\circ}$. For $\ge37.1~\mu$m, we conclude that any extended emission due to NLR dust is within the FWHM of the SOFIA instruments.

\textbf{NGC 4258}.
NGC 4258 was not detected but we include the data here since it is part of the sample. It has been observed and analyzed in the N-band with Gemini/Michelle by \citet{Mason2012}. These authors found a compact nucleus that is marginally resolved at 10 $\mu$m (FWHM $\sim0.5\arcsec$). 

\textbf{NGC 4388}.
NGC 4388 is an edge-on spiral that shows the most interesting mid- to far-IR morphology in this study. Notably, in the 30 - 40 $\mu$m FORCAST images of NGC 4388, extension can be seen in the NE to SW direction at PA$\sim$ 40 $^{\circ}$ \cite[see also][]{F19}, coincident with the NLR. This emission is seen on smaller scales at shorter wavelengths \citep{Asmus2016,GBernete2016}. The 53 $\mu$m image decreases in intensity and does not show a strong central core of emission as in the $31.5-37.1$ wavelength range. However, at longer wavelengths ($89-214~\mu$m), host galaxy emission clearly dominates the images in the east-west direction at PA$\sim$ 90$^{\circ}$.  

\textbf{NGC 4941}.
NGC 4941 is a low-luminosity AGN that appears here as a faint point-like source in the $31.5~\mu$m and $37.1~\mu$m images, but brighter at 53 and 89 $\mu$m. Subarcsecond resolution N-band imaging on VLT/VISIR \citep{Asmus2011} showed no significant extended MIR sources outside of the nucleus. 

\textbf{NGC 5506} 
NGC 5506 appears as a bright point source in both the 31.5 and 37.1 $\mu$m filters. While the nucleus is unresolved, extended MIR emission has been detected up to a few arcseconds to the northeast at 11.9 $\mu$m \citep{Raban2008}. Extended emission in the north-south direction was detected in the N-band out to $\sim$560 pc, while faint extended emission towards the east in the Q-band was also detected \citep{GBernete2016}.  However, the PSF-subtracted 12.27 $\mu$m 2"$\times$2" VLT/VISIR  image of \citet{AH2021} shows that the PA of extended emission varies from 30$^{\circ}$ in the central $\sim$0.5" to nearly 90$^{\circ}$ in the outer regions.

\textbf{NGC 7465}.
The 31.5 $\mu$m FORCAST image appears faint with a 3$\sigma$ upper-limit in the 37.1 $\mu$m image. The 53 and 89 $\mu$m HAWC+ images here appear increasingly brighter, albeit as point-like sources. Cold molecular gas observations \citep{Young2021} reveal that NGC 7465 is quite gas-rich, possibly from a recent merger.

\textbf{NGC 7469}.
NGC 7469 appears as a very bright source in the 31.5 $\mu$m image with FWHM $\sim$ 4.3". After PSF subtraction, \citet{F16} found extended emission in the north-south direction. This AGN is known to have a circumnuclear ring of star formation at a radius of $\sim$ 480 pc \citep[$\sim$ 1.4";][]{RA2011} in 8.7 and 18.3 $\mu$m images taken on Gemini/T-ReCS. Recent JWST observations reveal prominent PAH emission, indicative of star formation, in the circumnuclear ring \citep{IGB2022,Zhang2023}. 

\textbf{NGC 7674}.
The previously published \citep{F16} FORCAST 31.5 $\mu$m image appears as a point-like source. \citet{sasmirala} found that the nucleus of NGC 7674 is extended at PA $\sim$ 125$^{\circ}$ at subarcsecond scale resolution, where the extension roughly aligns with the ionization cone. 


\section{Nuclear Flux Extraction} 
\label{sec:analysis}

We aim to construct well-sampled mid- to far-IR SEDs of the nuclear emission of AGN at scales of several arcseconds, depending on the PSF of the observation and possible extended emission. On these scales, we expect multiple dust sources (i.e. torus, star forming regions, dusty outflow), however disentangling these sources is beyond the scope of this imaging atlas. Because the images span a range of observing cycles, instruments, and observing modes, we analyzed each image individually. Of our sample, 17 objects appear visually as point sources. For these sources, we performed aperture photometry where the aperture size was set to be $2\times$ the FWHM at a given band. For objects that show host galaxy emission, we extract the central PSF to construct the SEDs as described below. We complement our SOFIA data with \textit{Herschel} imaging data (see Appendix \ref{herschel_images}) and use a similar analysis method to construct the full IR SEDs. 

\subsection{Extended Sources: 2D Gaussian Fitting}
\label{sec:fitting}

For sources with extended dust emission, we performed a two-component simultaneous fit to accurately model both the central source based on the PSF, and the host galaxy whose fit assumes an elongated 2D Gaussian profile. In order to supplement the SOFIA data for the full mid- to far-IR SEDs, we used a similar methodology with \textit{Herschel} images.

For the SOFIA images, the PSF used was based on the standard stars of the observing runs for each cycle as explained in Section \ref{observations}. However, the same analysis could not be performed on \textit{Herschel} images due to the threefold lobes associated with the instrument PSFs. To accommodate this, we compared three different PSF models to reproduce and fit the central source. Two of the PSFs were point source images while the third was an approximation of the theoretical instrumental PSF using a Gaussian profile. The fitting routine used four free parameters for the Gaussian profile: (1,2) the position in x and y of the PSF center according to the image center, (3) the amplitude of the PSF, (4) the fourth parameter was dependent on the PSF type used. For archival PSFs, this parameter represents the rotation angle that needs to be applied to the PSF to match the orientation of the image. For the  Gaussian PSFs, this fourth parameter represents a scaling factor to the width of the Gaussian compared to its ideal value for a perfect instrument ($1.22\times \lambda / D$). 

The galaxy background is defined by 7 parameters: (1,2) the 2D Gaussian’s center position (x$_{0}$ and y$_0$), (3,4) its width ($\sigma_{x}$ and $\sigma_{y}$) in both directions, (5) its amplitude, and (6) its orientation on the image ($\theta$). To these 6 parameters we added a constant background as a 7th free parameter. We combined these components and fit this simulated intensity map to the observed map using the 11 total free parameters (4 from the PSF and 7 from the 2D gaussian). We thus derived the parameters describing the best central source for our intensity maps, and then studied the properties extracted for the central source and removed it from the initial map to study the host galaxy itself. An example of this procedure is given in the Appendix in Figure \ref{2dgauss}.

\subsection{AGN and host galaxy contribution}
\label{sec:host}

\begin{figure*}
\caption{PSF-subtracted images of host galaxy backgrounds}
\label{host_galaxy}
\includegraphics[scale=0.4]{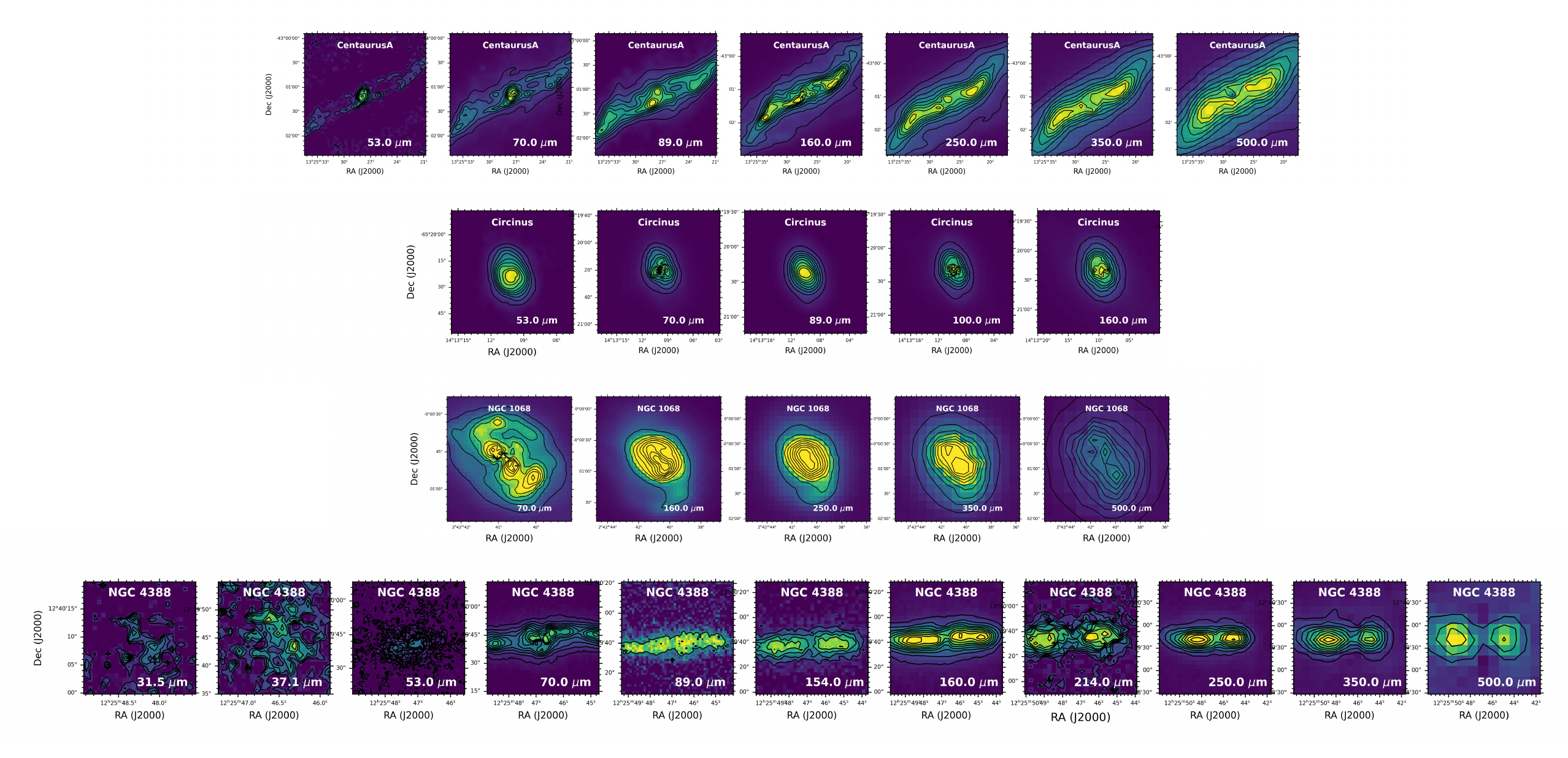}
\end{figure*}

While most SOFIA images were treated as point sources, Centaurus A, Circinus, NGC 1068, and NGC 4388 all had significant host galaxy contamination that needed to be subtracted from at least some of the images. Figure \ref{host_galaxy} shows the PSF subtracted images of these sources and Table \ref{psf_cont} gives the percentage of the contribution of the PSF to the total flux of the object.  In several other objects, the shorter wavelength ($\sim$ 30 - 100 $\mu$m) images did not show host galaxy contamination, but longer wavelength \textit{Herschel} images show the colder extended dust. Because this is an atlas of SOFIA images, we include objects with host galaxy contamination only in \textit{Herschel} images in Appendix \ref{herschel_host_galaxy} for completeness.

The 53 $\mu$m image of Centaurus A contains ($<5$\%) emission from extended sources, so we performed aperture photometry to account for the central emission. At wavelengths $\gtrsim 70 \mu$m, the host galaxy substantially ($\sim56-70$\%) contributes to the central AGN emission, so the extended emission was subtracted. The PSF of Centaurus A at these wavelengths contributes $\sim$ 35 \%. This can be interpreted as the nucleus having a relatively constant IR contribution, so the brightness of the nucleus coincides with IR brightness of the host galaxy.

The PSF contribution of the Circinus Galaxy decreases between 53 and 160 $\mu$m from 58 \% to  35 \%. At longer FIR wavelengths, the contribution of the PSF appears to be from the host galaxy and the fitting no longer provides information about the AGN. We interpret this as a decreasing IR contribution from the nucleus compared to the extended emission.

For the completeness of the SOFIA Atlas presented here, we used the 19-53 $\mu$m images of NGC 1068 from \citet{ELR2018}. These datasets were analyzed as described in that study and here we only present the results and images in that wavelength range. The study showed that the fractional contribution from star formation increases from 20 - 50 $\mu$m, while extended emission from 200 K dust decreases. Emission from the torus peaks in this range, a result which is in agreement with \citet{F16} who found that the turnover in torus emission occurs at wavelengths > 31.5 $\mu$m.  Extended emission is observed here at all wavelengths $\gtrsim$ 70 $\mu$m arising from dust in the host galaxy and star formation regions.

For NGC 4388, we show the results of PSF subtraction at all wavelengths, but only use the results in wavelengths > 40 $\mu$m for the SED. In the 30 - 40 $\mu$m range, the NE to SW extension is clear in the PSF-subtracted images. However, almost all of the extended emission lies within the FWHM of the observation; the FWHM of these images are only $\sim$ 10\% greater than the FWHM of the PSF. Thus, while we show the PSF subtracted images of NGC 4388 here, for the SED we use the total 30 - 40 $\mu$m fluxes which encompass the apparent extended emission due to the NLR. The change in the extended emission source and morphology between 40 and 70 $\mu$m is clear in the PSF-subtracted images (Figure \ref{host_galaxy}). The host galaxy clearly dominates the extended emission in the FIR while the NLR region dominates the extended emission in MIR wavelengths. The 53 $\mu$m HAWC+ image appears to show the transition between dominant extended sources. The contribution of the PSF in the images of NGC 4388 is $\sim$ 60 \% in the 30 - 40 $\mu$m range, where the extended emission is in the NE to SW direction. The contribution then decreases drastically to $\sim$ 20 \%. This reflects the turnover in extended emission seen in the images in Figure \ref{image4}. The contribution of the PSF returns to $\sim$ 70 \% between 70 - 100 $\mu$m, which suggests two separate but significant IR emission sources.

\begin{table*}

\caption{Contribution of the PSF to the host galaxy extended emission}
\centering
\label{psf_cont}
\footnotesize{
\begin{tabular}{cccccccccccccc}
\hline

 & \multicolumn{12}{c}{Wavelength ($\mu$m)} \\ \cline{2-14}
   & 19.7 &  31.5 & 37.1   &  53 & 70& 89 & 100 & 155 & 160 & 215 & 250   &350 & 500   \\ 
    \hline
Object & \multicolumn{12}{c}{Unresolved emission contribution (\%)} \\
\hline

Centaurus A & & & & \dots &   30$\pm$1 & 34$\pm$1 &  &  & 31$\pm$1 &  & 36$\pm$2 & 37$\pm$2 & 44$\pm$2 \\
Circinus   & & & & 58$\pm$0.2 & 48$\pm$1 & 35$\pm$0.3 & 40$\pm$1 & &        33$\pm$1 &  &  &  &  \\
NGC 1068 &36$\pm$4 &  49$\pm$5  & 51$\pm$8 & 33$\pm$15 & 56$\pm$2 & & & & 13$\pm$2  &  & 13 $\pm$ 1 & 35$\pm$1 & 54$\pm$1\\
NGC 4388 & & 62$\pm$16 & 56$\pm$12 & 19$\pm$2 & 64$\pm$1 &                71$\pm$2 & & 47$\pm$2 & 29$\pm$1 & 38$\pm$3 & 
    31$\pm$4 & 38$\pm$6 & 44$\pm$5 \\
    
    \hline

\end{tabular}}

\end{table*}

\section{Spectral Energy Distributions}
\label{sec:SEDs}

Tables \ref{fluxes} (SOFIA) and \ref{h_flux} (\textit{Herschel}) give the nuclear fluxes of the AGN in our sample along with their associated errors in units of Jy. The sources of uncertainty here are the instrument calibration, sky background, and the 2D gaussian fitting, where applicable. We estimate FORCAST and HAWC+ errors at $\sim$ 10\%. We use PACS instrument errors at 5\%\footnote{PACS: \url{https://www.cosmos.esa.int/web/herschel/pacs-overview}} and SPIRE instrument errors as 5.5\%\footnote{SPIRE: \url{https://www.cosmos.esa.int/web/herschel/spire-overview}}. The uncertainty due to sky background is determined on an individual basis, but averages $\sim$ 5\%. The average uncertainty due to the 2-D gaussian fitting is $\sim$ 1.5\%. We add these uncertainties in quadrature for the error bar estimation.

\begin{table*}
\centering

\caption{Nuclear fluxes for SOFIA/FORCAST and HAWC+ images. }

\begin{tabular}{cccccccc}
\hline
			&			&		&	Wavelength ($\mu$m)	&				&				&		&		Ref		\\
\hline
		        &31.5		&37.1	&	 53				& 89				&155				&	215	&				\\
\hline
Object		&			&		&	Flux (Jy)			&				&				&		&				\\		        

\hline	
Centaurus A	&			&		&	10.7	$\pm$ 1.1	& 18.2$\pm$2.1	&				&		&		This work		\\
			
Circinus		&			&		&	85.8	$\pm$ 8.7		&	84.7$\pm$8.52	&				&	51.5$\pm$5.2	&		This work		\\

MCG-5-23-16	&1.8$\pm$0.2&		&	1.7$\pm$ 0.2		&	0.7$\pm$0.1		&0.5$\pm$0.1	&		&	a, This work		\\
			
Mrk 3		&2.9$\pm$0.3	&3.0$\pm$0.3 &3.1$\pm$0.4		& 2.7$\pm$0.3	&2.1$\pm$0.2	&		&	b, This work		\\

Mrk 231		&			&		&					& 22.9$\pm$2.3	&				&		& 	This work			\\

Mrk 573		& 0.7 $\pm$0.1&		0.8 $\pm$0.1&		&				&				&		&		b, This work		\\

NGC 1068*	& 28.8$\pm$1.8 & 29.7$\pm$2.5 & 23.8$\pm$4.8	&			&				&		&		c,d \\

NGC 1275	& 4.0$\pm$ 0.4	& 5.0$\pm$0.8	&				&	5.3 $\pm$0.2		&				&		&		b, This work		\\

NGC 2110		& 1.3$\pm$0.2 & 			& 3.7$\pm$ 0.5	&  4.4 $\pm$ 0.5	& 4.5 $\pm$0.5	& 2.2$\pm$0.2 &	a, This work		\\

NGC 2273	& 1.9$\pm$0.2	& 2.7$\pm$0.3 & 4.3 $\pm$ 0.5	& 6.5 $\pm$ 0.7	& 5.4 $\pm$0.6	& 3.8$\pm$0.4 &	a, This work		\\

NGC 2992	& 0.9$\pm$0.1 & 			&				&				&				&				&	a	\\

NGC 3081	& 1.0$\pm$0.1 & 1.4$\pm$0.2 &				&				&				&				&	a,b	\\

NGC 3227	& 2.3$\pm$0.3 & 2.8$\pm$0.3 & 				&				&				&				&	a,b	\\

NGC 3281	& 2.7$\pm$0.3 &			& 5.3$\pm$ 0.6	& 6.4$\pm$0.7	& 4.6$\pm$0.5 & 2.5$\pm$0.3	&	a, This work	\\

NGC 4151	& 3.9$\pm$0.4	 & 4.5$\pm$0.3 & 4.9$\pm$0.5  & 4.1$\pm$ 0.4	&				&				&	b, This work	\\

NGC 4258    &               &               &               &                   &               &               &   This work    \\

NGC 4388	& 3.0$\pm$0.4 & 3.2$\pm$0.3 & 0.7$\pm$0.1	& 3.3$\pm$ 0.4	& 3.4$\pm$0.4	& 2.7$\pm$0.3	&	a,b, This work	\\

NGC 4941	&0.29$\pm$0.04	&	0.5$\pm$0.1	&	&				&				&				&	This work	\\

NGC 5506	& 4.1$\pm$0.5 &	5.2$\pm$0.5	&		&	7.0$\pm$0.2			&				&				&	a, This work,e	\\

NGC 7465    &            &          &  1.8$\pm$0.2    &     4.9$\pm$0.5      &           &           &   This work    \\

NGC 7469	& 9.4$\pm$1.1 &			&				&				&				&				&	a	\\

NGC 7674	& 1.8$\pm$0.2 &			&				&				&				&				&	a	\\
\hline
\end{tabular}

\label{fluxes}

\begin{tablenotes}
\item \textsc{References}: a) \citet{F16}, b) \citet{F19}, c) \citet{ELR2018}, d) \citet{ELR2020}, e) \citet{ELR2022a}. 
*\citet{ELR2018} measured the flux of NGC 1068 at 19.7 $\mu$m to be 22.0$\pm$1.4.
\end{tablenotes}

\end{table*}

\begin{table*}
\centering
\caption{Nuclear fluxes for \textit{Herschel}/PACS and SPIRE images.}
\begin{tabular}{ccccccc}
\hline
    &   &   &   Wavelength ($\mu$m) &   &    &      \\
\hline
    &   70  &   100 &   160 &   250 &   350 &   500     \\
\hline
Object  &   &   &   Flux (Jy)   &   &   &   \\
\hline
Centaurus A &   15.6$\pm$1.0   &   &   19.9$\pm$2.0   &   9.9$\pm$1.0 &  6.9$\pm$0.5 & 5.8$\pm$0.4 \\
Circinus    &   99.0$\pm$5.3 & 63.5$\pm$3.3 & 47.2$\pm$2.4 & 20.1$\pm$1.1 & 18.9$\pm$1.2 & 6.5$\pm$0.4 \\
MCG-5-23-16 & 1.1$\pm$0.1 &   & 0.34$\pm$0.05 & 0.11$\pm$0.02 & 0.05$\pm$0.01 & 0.02$\pm$0.005 \\
Mrk 3 & 2.9$\pm$0.2 & 2.6$\pm$ 0.1 & 1.9$\pm$0.1 & 0.9$\pm$0.1 & 0.6$\pm$0.1 & 0.34$\pm$0.09 \\
Mrk 231 & 31.2$\pm$1.6 & 26.4$\pm$1.4 & 13.9$\pm$0.7 & 5.2$\pm$0.3 & 1.8$\pm$0.1 & 0.49$\pm$0.03 \\
Mrk 573 & 1.02$\pm$0.05 & 1.05$\pm$ 0.61& 0.94$\pm$0.06 & 0.44$\pm$0.03 & 0.23$\pm$0.02 & 0.08$\pm$0.02 \\
NGC 1068 & 101.7$\pm$2.2 &  & 142.2$\pm$0.5 & 69.7$\pm$0.5 & 27.4$\pm$0.4 & 8.8$\pm$0.3 \\
NGC 1275 & 6.8$\pm$0.4 & 6.7$\pm$0.4 & 5.5$\pm$0.3 & 3.2$\pm$0.2 & 2.7$\pm$0.2 & 2.4$\pm$0.1 \\
NGC 2110 & 4.3$\pm$0.2 & 5.0$\pm$0.3 & 4.2$\pm$0.2 & 1.7$\pm$0.1 & 0.57$\pm$0.09 & 0.20$\pm$0.05 \\
NGC 2273 & 5.2$\pm$0.3 & 6.0$\pm$0.3 & 5.8$\pm$0.3 & 2.3$\pm$0.1 & 0.93$\pm$0.06 & 0.19$\pm$0.02 \\
NGC 2992 & 2.4$\pm$0.1 & 3.3$\pm$0.2 & 3.5$\pm$0.2 & 1.9$\pm$0.1 & 0.74$\pm$0.04 & 0.54$\pm$0.04 \\
NGC 3081 & 1.9$\pm$0.1 & 2.1$\pm$0.1 & 1.6$\pm$0.1 & 0.67$\pm$0.04 & 0.24$\pm$0.02 & 0.11$\pm$0.02 \\
NGC 3227 & 6.3$\pm$0.3 & 8.3$\pm$0.4 & 4.8$\pm$0.3 & 2.1$\pm$0.1 & 0.68$\pm$0.04 & 0.16$\pm$0.01 \\
NGC 3281 & 6.4$\pm$0.3 & 6.3$\pm$0.3 & 4.5$\pm$0.2 & 1.3$\pm$0.1 & 0.38$\pm$0.03 & 0.25$\pm$0.03 \\
NGC 4151 & 4.4$\pm$0.2 & 3.1$\pm$0.2 & 1.5$\pm$0.1 & 0.40$\pm$0.02 & 0.06$\pm$0.005 & 0.0007$\pm$7e${^-5}$ \\
NGC 4258 &  1.1$\pm$0.2 & 4.1$\pm$0.3   & 4.7$\pm$0.5   & 3.3$\pm$0.3   & 2.5$\pm$0.3   & 0.9 $\pm$0.1   \\
NGC 4388 & 2.9$\pm$0.2 &  & 3.5$\pm$0.2 & 1.1$\pm$0.1 & 0.91$\pm$0.08 & 0.49$\pm$0.04 \\
NGC 4941 &  0.80$\pm$0.04 &     & 0.70$\pm$0.08 & 0.12$\pm$0.01 & 0.015$\pm$0.001 & 0.05$\pm$0.006 \\
NGC 5506 & 6.8$\pm$0.4 & 6.4$\pm$0.3 & 3.9$\pm$0.2 & 1.2$\pm$0.1 & 0.53$\pm$0.03 & 0.26$\pm$0.03 \\
NGC 7465 & 3.6$\pm$0.2 & 4.9$\pm$0.3 & 4.1$\pm$0.2 & 1.8$\pm$0.1 & 0.89$\pm$0.07 & 0.42$\pm$0.04 \\
NGC 7469 & 22.8$\pm$1.2 & 25.9$\pm$1.3 & 18.3$\pm$0.9 & 7.5$\pm$0.4 & 2.9$\pm$0.2 & 0.79$\pm$0.05 \\
NGC 7674 & 3.9$\pm$0.2 & 4.5$\pm$0.2 & 4.9$\pm$0.3 & 2.7$\pm$0.2 & 1.2$\pm$0.1 & 0.44$\pm$0.04 \\
\hline
\label{h_flux}
\end{tabular}
\end{table*}


\begin{figure*}
\centering
\includegraphics[scale=0.7]{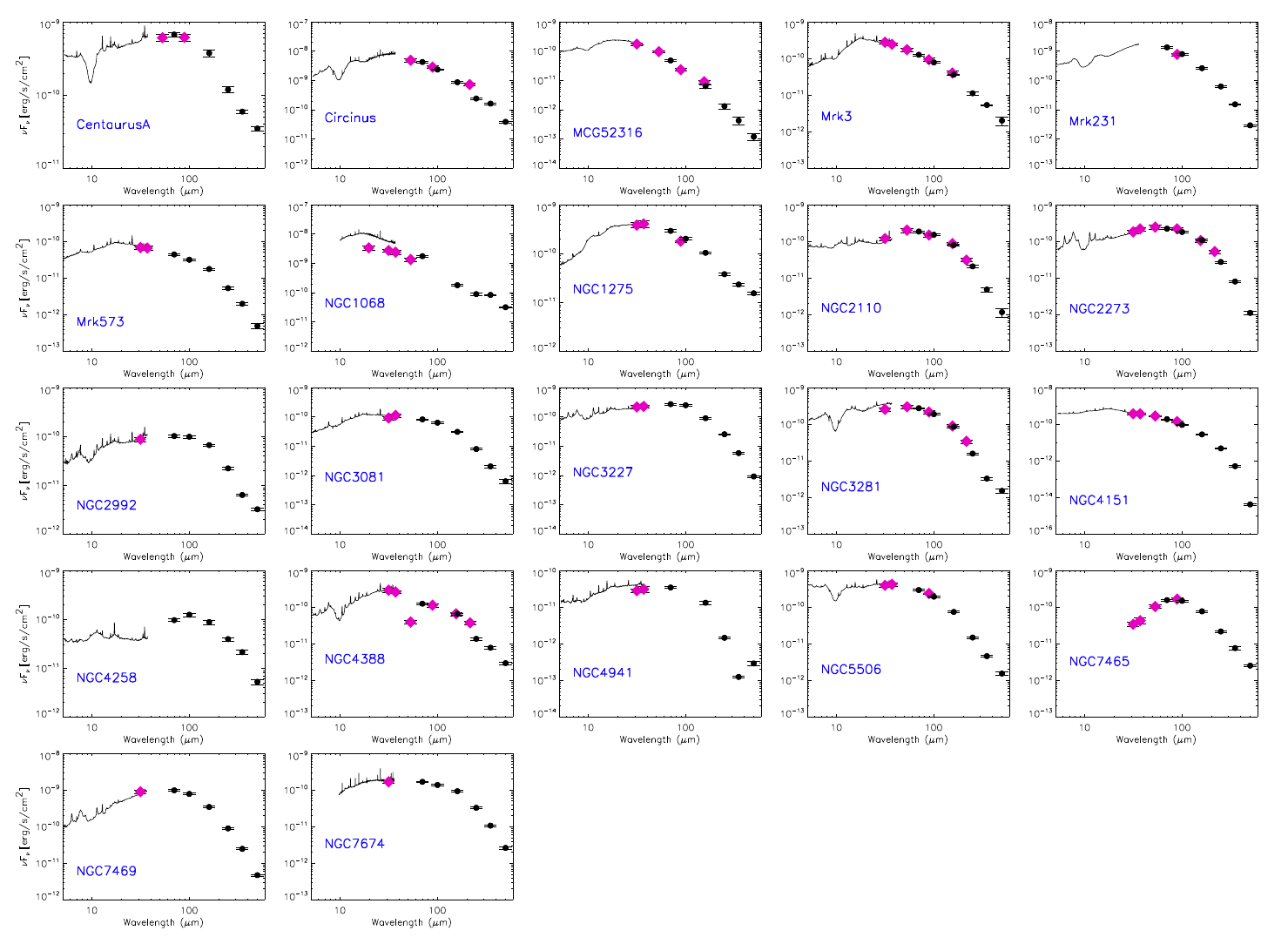}
\caption{Mid- to far-IR SEDs of the several-arcsecond-scale nuclear fluxes in our sample of AGN. Pink diamonds represent SOFIA observations while black dots represent complementary Herschel observations. The solid black lines correspond to \textit{Spitzer} spectra.}
\label{seds}
\end{figure*}

The nuclear SEDs are shown in Figure \ref{seds} in $\nu$F$_{\nu}$. The pink diamonds represent SOFIA observations while the black circles represent the complementary \textit{Herschel} data. We obtained \textit{Spitzer}/IRS spectra from the \textit{Sptizer}/CASSIS database \citep{Lebouteiller2011} for 21 of the 22 objects in our sample (solid black line). There was no spectrum available for NGC 7465.  Low-resolution spectra (R $\sim$ 100) were obtained for 18 of the objects, while moderate resolution spectra (R $\sim$ 600) were available for Circinus, NGC 1068, and NGC 7674. This dataset provides the most completed SED coverage available between 30 - 500 $\mu$m. Decomposing the SEDs in this sample is outside the scope of this manuscript, as we are presenting an imaging atlas. Here, we provide the main results and features of the SEDs of these objects.

 The morphological changes seen in the extended emission source in Figure \ref{image4} for NGC 4388 are reflected in the SED at 53 $\mu$m, where there is a marked decline in the SED. The drastic decrease seems to be due the change of dominant emitting sources. The extended emission at wavelengths $\lesssim$ 40 $\mu$m is due to dust in the direction of the radio axis, and the extended emission at wavelengths $\gtrsim$ 50 $\mu$m is due to the host galaxy.

The wavelength of peak emission can give insight to the primary processes that drive MIR emission. The peak wavelength, determined by the highest flux from photometry and spectroscopy, ranges from $18$ to $100$ $\mu$m in $\nu$F$_\nu$ with an average of $\sim$ 40 $\mu$m. This average only includes the peak in continuum values and does not take into account fine structure lines. Most ($73$\%; 11 out of 15) Seyfert 2 have a peak emission at wavelengths $\lesssim 40 \mu$m. The SEDs of MCG-5-23-16, Mrk3, Mrk 573, NGC 1068, NGC 3081, and NGC 4151 peak at $\sim$ 18 - 20 $\mu$m. This is in agreement with the correlation peak between the hard-X-rays and the mid-IR for Type 1 AGN in \citet{IGB2017}. The peak wavelengths in $F_{\nu}$ (Jy) range $\sim$ 20 - 160 $\mu$m, with an average $\sim$ 93 $\mu$m. NGC 1068 is the only AGN to peak at the same wavelength in both sets of units. 

 The $Spitzer$ spectrum for NGC 1068 does not align with the SOFIA photometry because of the extensive PSF subtraction that we performed in the photometry that was not accounted for in the spectroscopy. This is the only object that not only has overlapping 20 - 40 $\mu$m $Spitzer$ and SOFIA data, but also that has had the background emission subtracted at these wavelengths.

\subsection{Luminosity and Peak Wavelength}

\begin{figure}
\centering
\includegraphics[width=\columnwidth]{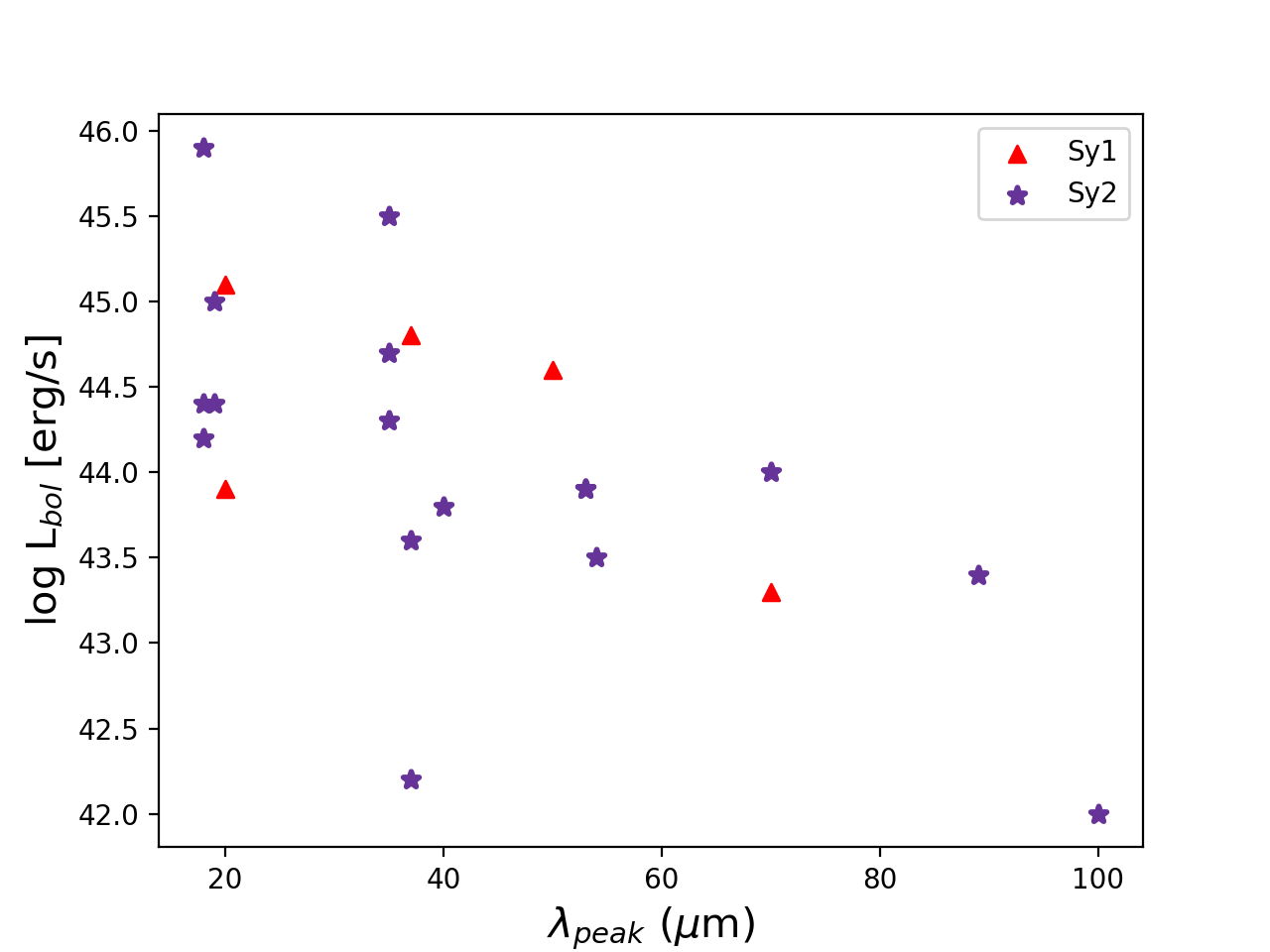}
\caption{Bolometric luminosity vs the peak wavelength of the SED for both Sy1 (red triangles) and Sy2 (purple stars). }
\label{plot_lbol}
\end{figure}

To test whether the peak wavelength is a function of luminosity, we plot L$_{bol}$ vs $\lambda_{peak}$. Figure \ref{plot_lbol} shows the bolometric luminosities of the AGN plotted against the peak wavelength in the SEDs for both Sy1s, shown as red triangles, and Sy2s, shown as purple stars. The correlation coefficient between the luminosity and peak wavelength is $|R| \sim$ 0.63 with statistical significance $p = 0.0015$. While it is argued that $|R|$ of 0.6 - 0.7 \textit{may} show moderate to strong correlation \citep[see Section 3.2 in][]{MSV2013}, a $p$-value $\le$ 0.05 is generally accepted as statistically significant. The data suggests that higher luminosity objects have SEDs that peak at shorter wavelengths, which indicates the presence of a hot dust component in the vicinity of the AGN.

\subsection{Mid- to Far-IR Colors}

\begin{figure}
\includegraphics[scale=0.55]{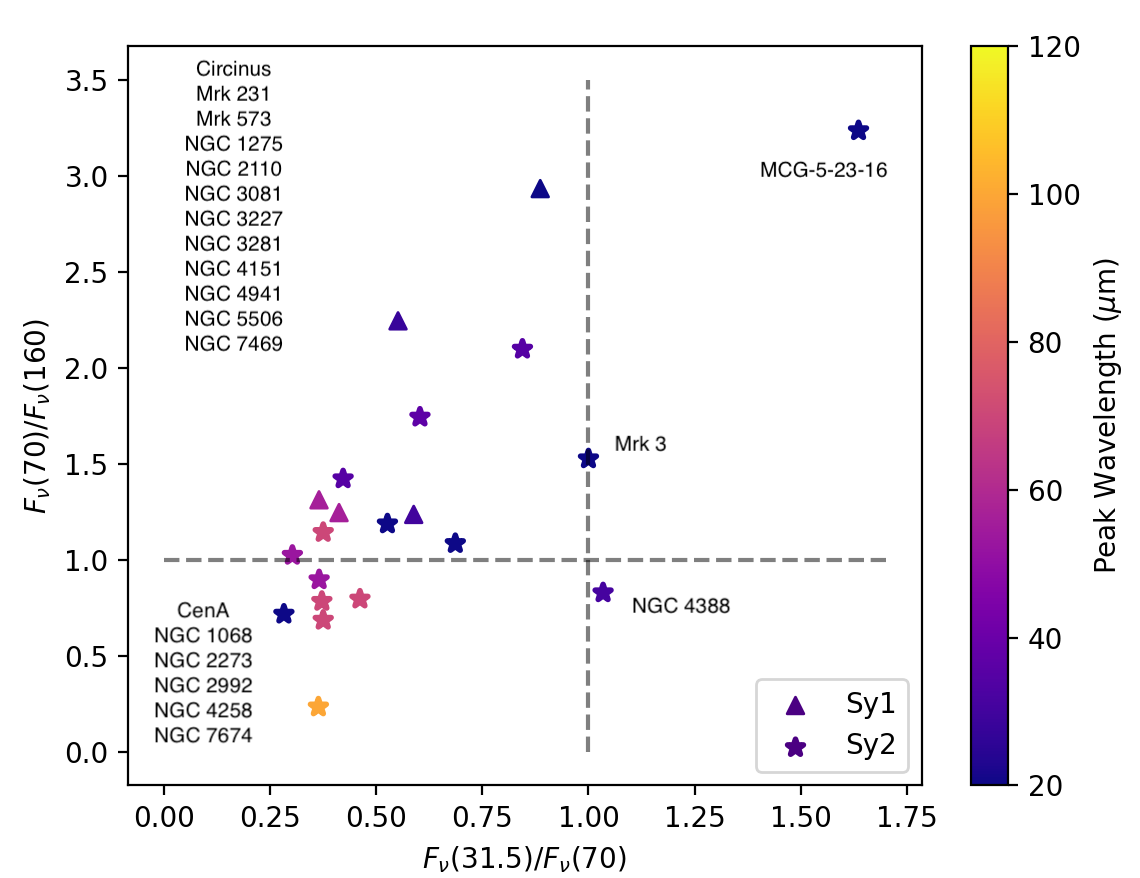}
\caption{Color diagram of 21 of the 22 AGN in our sample. Sy 1 are represented by triangles while Sy 2 are stars. The scale on the right shows peak wavelength by color. }
\label{color_plots}
\end{figure}

The ratio of F$_{\nu}$(70)/F$_{\nu}$(160) has been used as a proxy for dust temperature \citep{Melendez2014,GG2016}, where the ratio is higher for dust heated by the AGN and lower for dust heated by star formation. Here we perform this analysis using the ratio  F$_{\nu}$(31)/F$_{\nu}$(70) by using the 31.5 $\mu$m SOFIA data in our atlas. For objects that do not have data in the 31.5 $\mu$m filter, we supplement that with data from the \textit{Spitzer}/IRS continuum. NGC 7465 did not have 31.5 $\mu$m flux data, nor did it have \textit{Spitzer} data so we leave that object out of this analysis. Using the fluxes in Table \ref{fluxes}, we plot a color-color diagram in F$_{\nu}$ in Figure \ref{color_plots}. This figure also visually shows the peak wavelength from the SEDs (in $\nu F_{\nu}$). Longer peak wavelengths tend to cluster at F$_{\nu}$(70)/F$_{\nu}$(160) $\sim$ 1 and F$_{\nu}$(31)/F$_{\nu}$(70) between 0.25 - 0.5.

In this sample we find an average F$_{\nu}$(70)/F$_{\nu}$(160) ratio of $1.4\pm 0.7$. Previous studies \citep{Melendez2014,GG2016} with larger sample sizes (313 and 33, respectively) have found an average ratio of $\sim$0.8, albeit the data was analyzed using independent methods. This suggests a higher amount of AGN heated dust in our sample. 

We find that 18 objects have F$_{\nu}$(31)/F$_{\nu}$(70) <1, with an average F$_{\nu}$(31)/F$_{\nu}$(70) of $0.6\pm 0.3$. The only object with both F$_{\nu}$(70)/F$_{\nu}$(160) and F$_{\nu}$(31)/F$_{\nu}$(70) >1 is MCG-5-23-16, and an SED that peaks $\sim$ 20 $\mu$m. This object may be the most AGN dominated source in our sample. The other objects that show a peak at $\sim$ 20 $\mu$m in $\nu$F$_{\nu}$ still show F$_{\nu}$(31)/F$_{\nu}$(70) <1. Only one object, NGC 4388, shows F$_{\nu}$(31)/F$_{\nu}$(70) >1 while  F$_{\nu}$(70)/F$_{\nu}$(160) <1. This reflects the change in extended emission seen in Figure \ref{image4}. 

Half (11) of the objects in the sample show ratios F$_{\nu}$(31)/F$_{\nu}$(70) <1 while F$_{\nu}$(70)/F$_{\nu}$(160) >1. These objects (Circinus, Mrk 231, Mrk 573, NGC 1275, NGC 2110, NGC 3081, NGC 3227, NGC 3281, NGC 4151, NGC 4941, NGC 5506, NGC 7469) are likely AGN dominated.  Six objects show F$_{\nu}$(31)/F$_{\nu}$(70) <1 and F$_{\nu}$(70)/F$_{\nu}$(160) <1, meaning that their SEDs peak at longer wavelengths. The emission from these objects (Centaurus A, NGC 1068, NGC 2273, NGC 2992, NGC 4258, NGC 7674) are likely dominated by star formation.

\section{Conclusions}
\label{conclusions}

We have presented a SOFIA atlas of nearby AGN in the 20 - 215 $\mu$m wavelength range using FORCAST and HAWC+. We have released 69 observations of which 41 are newly published and 28 have been previously published \citep{F16,F19,ELR2018,ELR2022d}. From these observations, NGC 4388 shows the most dramatic visual change in emission morphology. The 30 - 40 $\mu$m images show a NE to SW dusty extension associated with the NLR, while the > 50 $\mu$m images show a East to West dusty emission associated with the plane of the host galaxy. Our observations show that $<10\arcsec$ resolution 30 - 70 $\mu$m  observations are crucial to disentangle the emitting contribution from AGN and host galaxy.

We measured arcsecond scale unresolved nuclear fluxes in order to construct SEDs of the objects in our sample. We included complementary \textit{Herschel} data to cover up to 500 $\mu$m. For point sources we used aperture photometry to determine the flux.  For extended sources we used a 2D gaussian fitting method to extract the central unresolved source(s) of emission from the galaxy background. For this method, the PSF is scaled to represent the central emission while a 2D gaussian represents host galaxy or background emission. Based on the SEDs, we make the following conclusions: \\
- There is a sharp drop in the SED of NGC 4388 that corresponds to the wavelength where the angle of extended emission transitions from NE/SW (NLR) to E/W (host galaxy).\\
- The average peak of the SEDs is 40 $\mu$m in $\nu$F$_{\nu}$, spanning a range of [20,100] $\mu$m.\\
- The peak wavelength of the SED appears to be a function of AGN luminosity, where higher luminosity objects peak at shorter wavelengths.\\
- MCG-5-23-16 is the only object whose color diagram shows both F$_{\nu}$(31)/F$_{\nu}$(70) and F$_{\nu}$(70)/F$_{\nu}$(160) >1, which may indicate an AGN dominated source. \\
- Half of the objects in the sample have flux ratios which suggest that the SED is dominated by AGN heated dust, while six objects show ratios consistent with heating by SF. \\

In future studies, we will combine data from this atlas with incoming data from JWST to update our IR datasets with the latest and highest resolution data available. Newly obtained JWST/MIRI observations will provide new higher angular resolution data for some of the sources in the wavelength range 5 - 28 $\mu$m. 

\section*{Acknowledgments}
We acknowledge Dr. Lucas Grosset for his effort in subtracting the image backgrounds. E.L.-R. is supported by the NASA/DLR Stratospheric Observatory for Infrared Astronomy (SOFIA) under the 08\_0012 Program. SOFIA is jointly operated by the Universities Space Research Association,Inc.(USRA), under NASA contract NNA17BF53C, and the Deutsches SOFIA Institut (DSI) under DLR contract 50OK0901 to the University of Stuttgart. E.L.-R. is supported by the NASA Astrophysics Decadal Survey Precursor Science (ADSPS) Program (NNH22ZDA001N-ADSPS) with ID 22-ADSPS22-0009 and agreement number 80NSSC23K1585. I.G.B. acknowledges support from STFC through grants ST/S000488/1 and ST/W000903/1. C.R. acknowledges support from Fondecyt Regular grant 1230345 and ANID BASAL project FB210003.

\software{\textsc{astropy} \citep{astropy:2022,astropy:2018,astropy:2013}; 
\textsc{SciPy} \citet{SciPy}}

\bibliographystyle{aasjournal}
\bibliography{atlas.bib}{}

\appendix

\section{\textit{Herschel} Images}
\label{herschel_images}

\begin{figure*}

\centering
\includegraphics[scale=0.9]{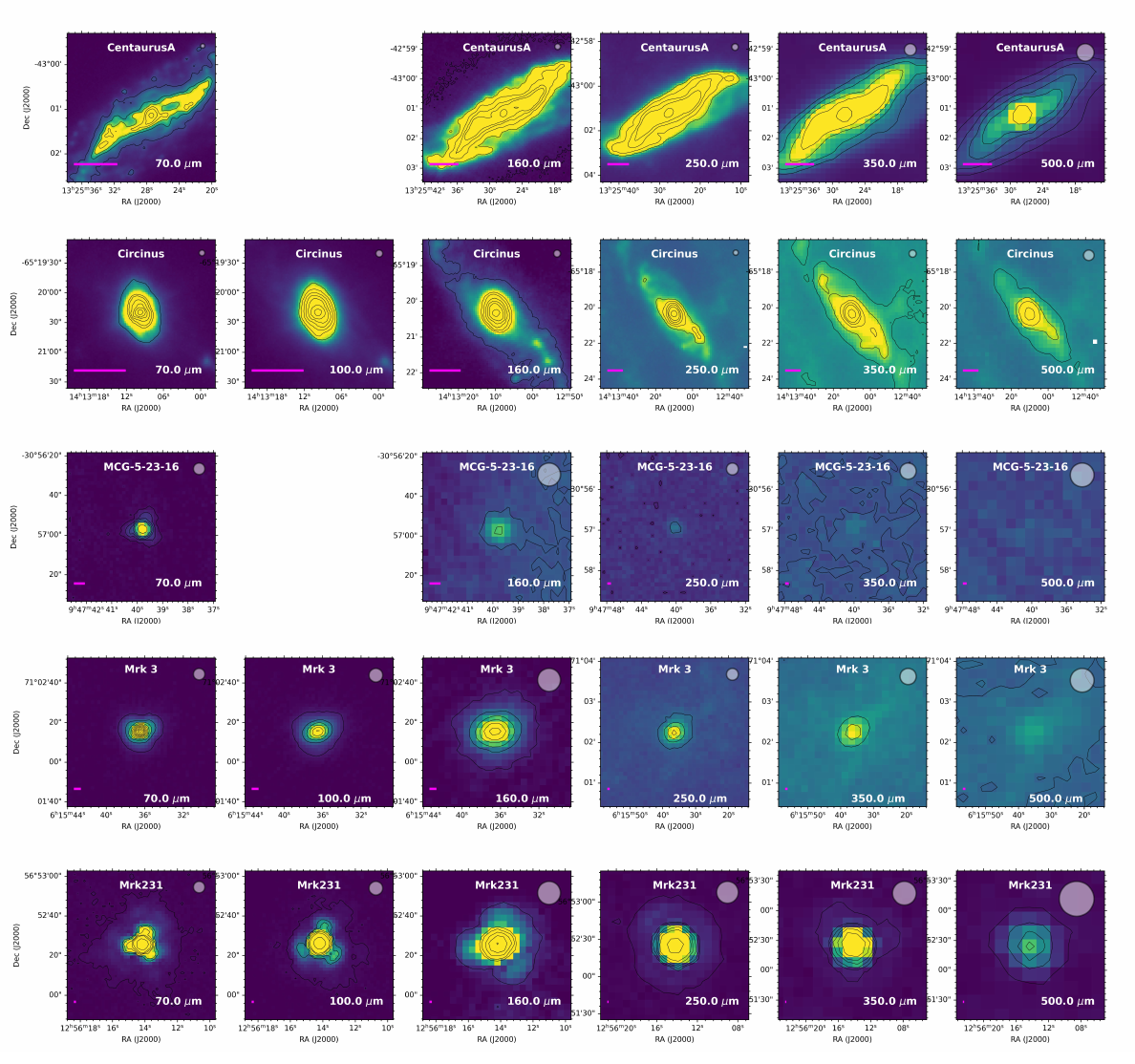}
\caption{\textit{Herschel} PACS and SPIRE FIR images. The white circle in the top right indicates the beam size of the observation, while the pink line in the bottom left indicates a distance of 1 kpc.}
\label{appendix:herschel1}
\end{figure*}

\begin{figure*}
\centering
\includegraphics[scale=0.9]{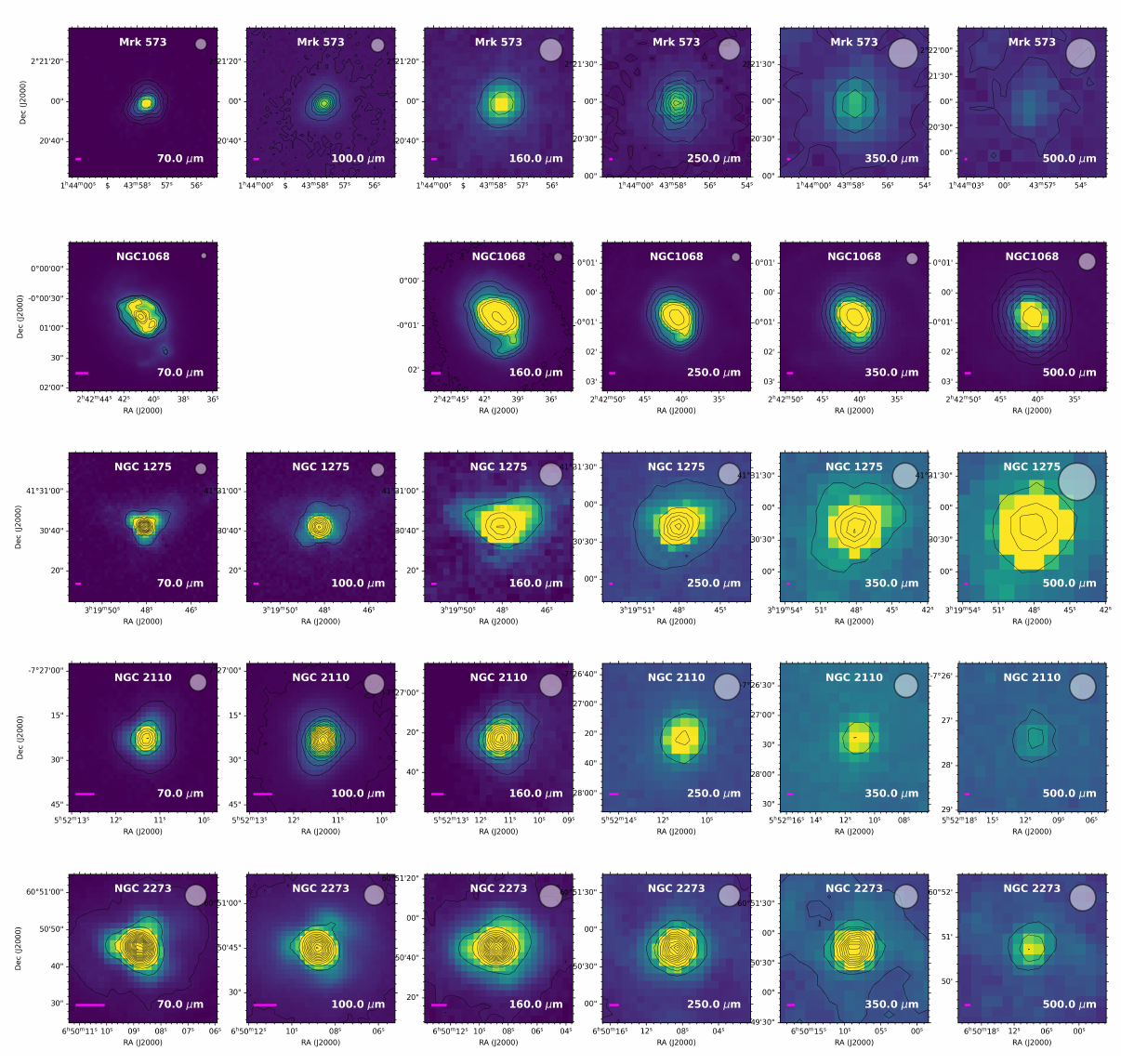}
\caption{\textit{Herschel} PACS and SPIRE FIR images. The white circle in the top right indicates the beam size of the observation, while the pink line in the bottom left indicates a distance of 1 kpc.}
\label{appendix:herschel2}
\end{figure*}

\begin{figure*}
\centering
\includegraphics[scale=0.9]{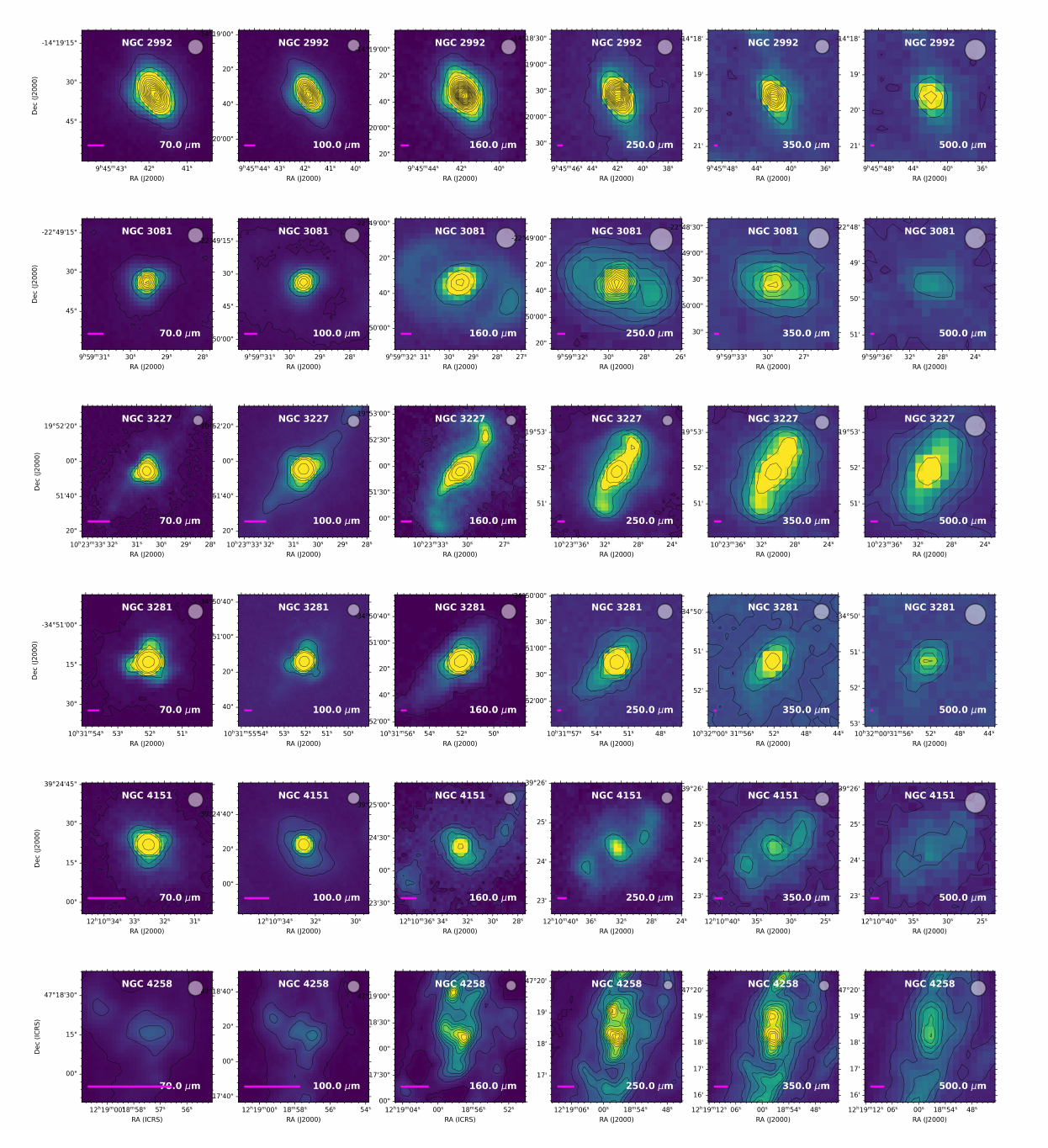}
\caption{\textit{Herschel} PACS and SPIRE FIR images. The white circle in the top right indicates the beam size of the observation, while the pink line in the bottom left indicates a distance of 1 kpc.}
\label{appendix:herschel3}
\end{figure*}

\begin{figure*}
\centering
\includegraphics[scale=0.9]{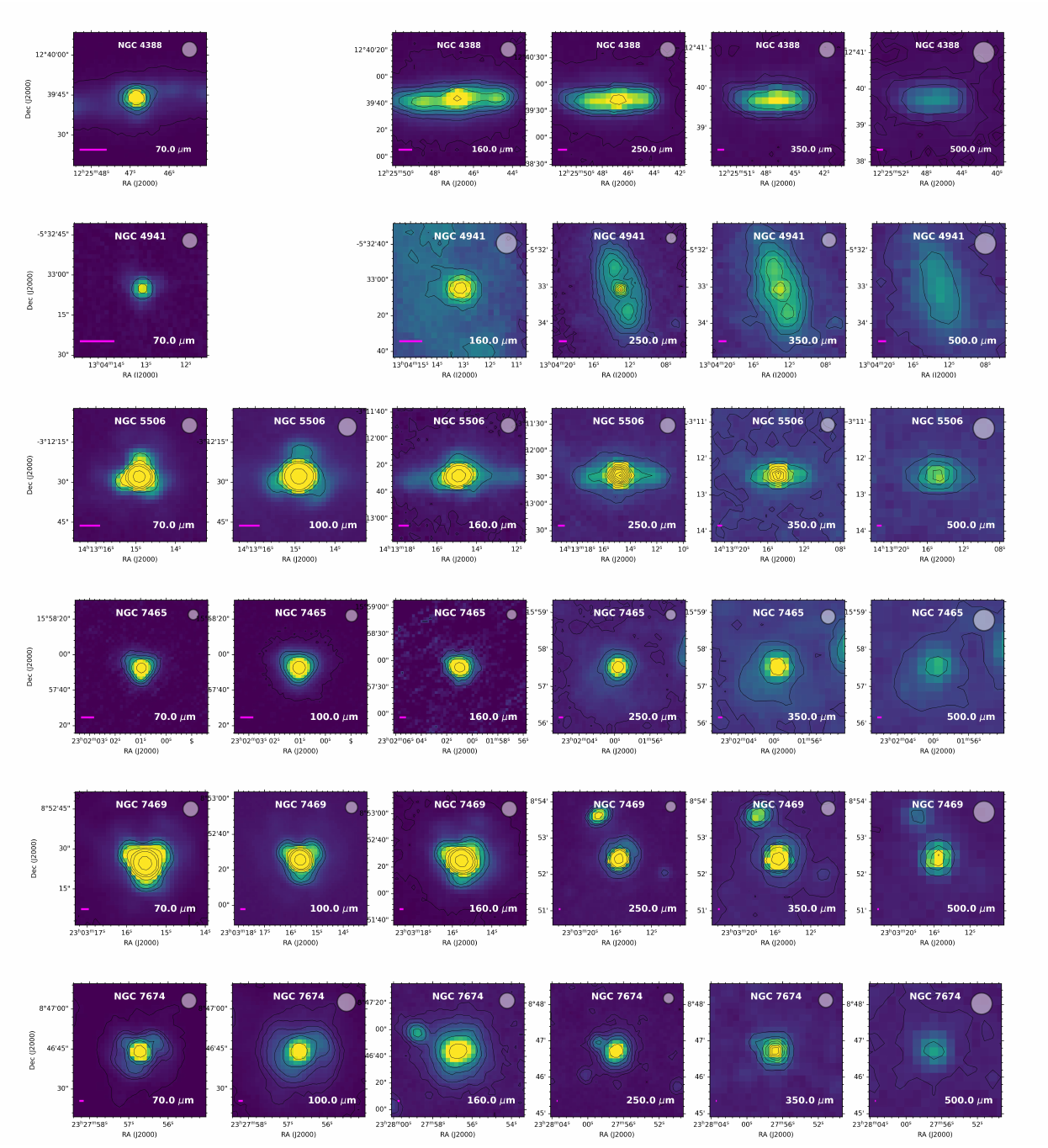}
\caption{\textit{Herschel} PACS and SPIRE FIR images. The white circle in the top right indicates the beam size of the observation, while the pink line in the bottom left indicates a distance of 1 kpc.}
\label{appendix:herschel4}
\end{figure*}

We used images from the \textit{Herschel} Archive to supplement SOFIA data in FIR wavelengths. Images from the PACS and SPIRE instruments covering the wavelength range 70 - 500 $\mu$m are shown in Figures \ref{appendix:herschel1}, \ref{appendix:herschel2}, \ref{appendix:herschel3}, and \ref{appendix:herschel4}. The white circle in the upper right indicates the beam size while the pink scale in the bottom left indicates a distance of 1 kpc. To extract nuclear fluxes on scales of several arcseconds, we use the methods described in Section \ref{sec:analysis}.  For point sources, we performed aperture photometry. For extended sources, we used the 2D gaussian routine outlined in Section \ref{sec:fitting}. 

\section{Background fitting}

\begin{figure*}

\includegraphics[scale=0.5]{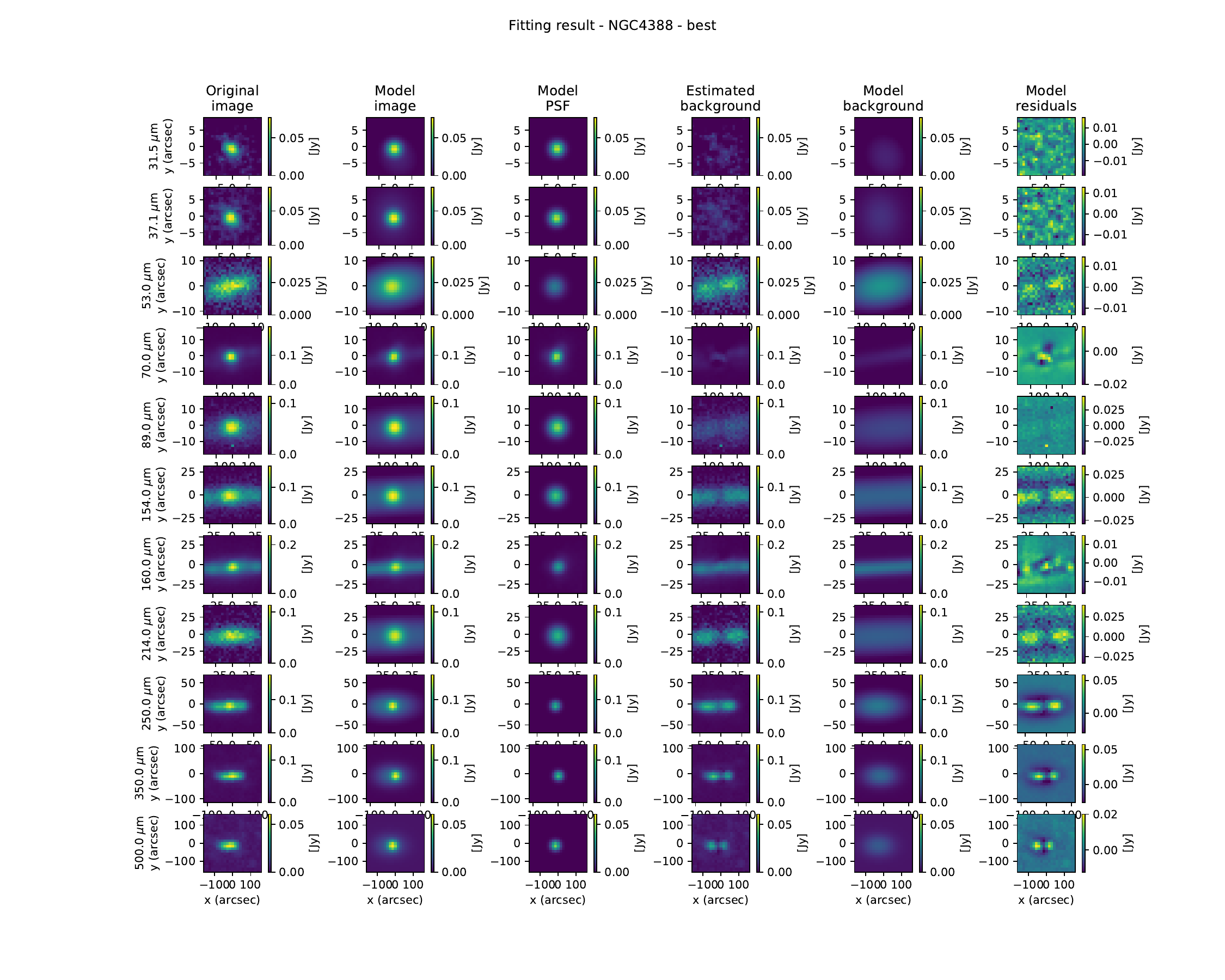}

\caption{Example of the 2D gaussian fitting routine explained in Section \ref{sec:fitting}}
\label{2dgauss}
\end{figure*}

Figure \ref{2dgauss} shows an example of the 2-D gaussian fitting we used to extract the central flux from the background of the host galaxy in wavelengths 30 - 500 $\mu$m. The object used here is NGC 4388. In the first column, the original observation image is shown. The second column shows a model of the image using a PSF (Column 3) combined with the model background. Column 4 shows a PSF-subtracted image (Column 1 - Column 3) used to make the model background (Column 5). Column 6 shows the PSF- and background- subtracted image of the residuals. 

\section{Host galaxy backgrounds - \textit{Herschel}}
\label{herschel_host_galaxy}

Here we show the results of PSF-subtracted images for objects in which only \textit{Herschel} data showed host galaxy contamination.

\begin{figure*}
\centering
\includegraphics[scale=0.5]{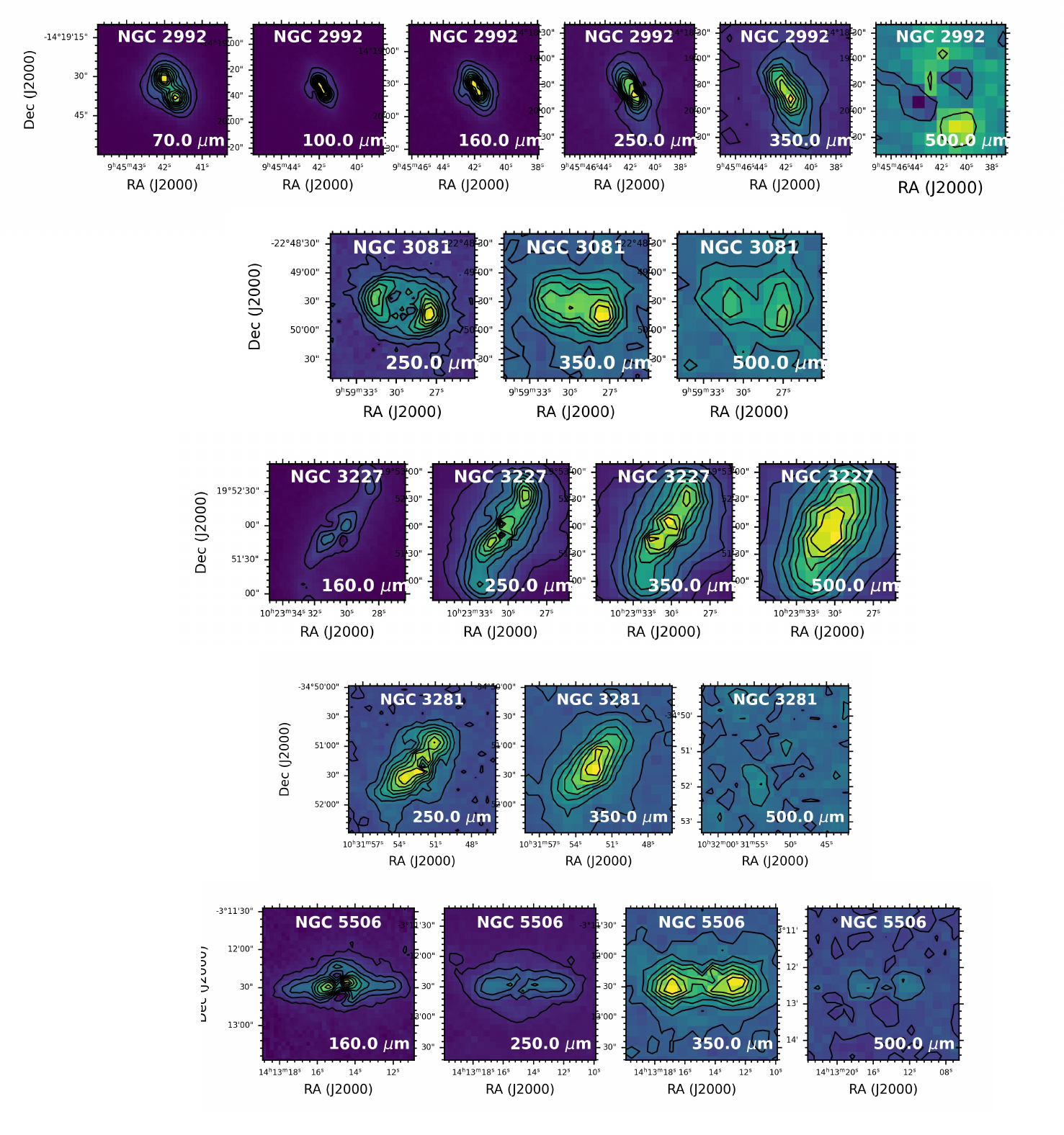}
\caption{PSF-subtracted host galaxy images from archive \textit{Herschel} data}
\end{figure*}

\end{document}